\documentclass[fleqn,usenatbib]{mnras}

\usepackage[T1]{fontenc}

\DeclareRobustCommand{\VAN}[3]{#2}
\let\VANthebibliography\thebibliography
\def\thebibliography{\DeclareRobustCommand{\VAN}[3]{##3}\VANthebibliography}

\usepackage{graphicx}	
\usepackage{amsmath}	
\usepackage{amssymb}	

\usepackage{newtxtext,newtxmath}

\title[Lyman-$\alpha$ transits]{Using Lyman-$\alpha$ transits to constrain models of atmospheric escape}

\author[E. Schreyer et al.]{
Ethan Schreyer,$^{1,3}$\thanks{E-mail: es2516@ic.ac.uk}
James. E. Owen,$^{1}$
R. O. Parke Loyd$^{2}$
and Ruth Murray-Clay$^{3}$
\\
$^{1}$Astrophysics Group, Imperial College London, Blackett Laboratory, Prince Consort Road, London SW7 2AZ, UK\\
$^{2}$Eureka Scientific, Inc., Oakland, CA 94602, USA\\
$^{3}$Department of Astronomy and Astrophysics, University of California, Santa Cruz, CA 95064, USA
}

\date{Accepted XXX. Received YYY; in original form ZZZ}

\pubyear{2024}

\begin{document}
\label{firstpage}
\pagerange{\pageref{firstpage}--\pageref{lastpage}}
\maketitle

\begin{abstract}
Lyman-$\alpha$ transits provide an opportunity to test models of atmospheric escape directly. However, translating observations into constraints on the properties of the escaping atmosphere is challenging. The major reason for this is that the observable parts of the outflow often comes from material outside the planet's Hill sphere, where the interaction between the planetary outflow and circumstellar environment is important. As a result, 3D models are required to match observations. Whilst 3D hydrodynamic simulations are able to match observational features qualitatively, they are too computationally expensive to perform a statistical retrieval of properties of the outflow. Here, we develop a model that determines the trajectory, ionization state and 3D geometry of the outflow as a function of its properties and system parameters.  We then couple this model to a ray tracing routine in order to produce synthetic transits. We demonstrate the validity of this approach, reproducing the trajectory of the outflows seen in 3D simulations. We illustrate the use of this model by performing a retrieval on the transit spectrum of GJ 436 b. The bound on planetary outflow velocity and mass loss rates are consistent with a photoevaporative wind. 

\end{abstract}




\section{Introduction}

Many of the discovered exoplanets reside close to their host stars. These planets are highly irradiated, causing their upper atmospheres to be extremely hot. The resulting pressure gradients can drive the gas to escape hydrodynamically. For small, low-mass planets ($1-4 \, {\rm R}_{\oplus}$, $\lesssim 20$~M$_\oplus$), the integrated effect of this mass loss can completely strip them of their primordial hydrogen/helium envelopes \citep{Valencia2010, Owen2013, Lopez2013, Chen2016}. This process can explain the observed lack of Neptune-sized planets at short orbital periods, the "hot Neptune desert" \citep{Szabo2011, Lundkvist2016}, and the observed sparsity of planets with radii $\sim 1.8R_{\oplus}$, referred to as "the radius gap" \citep{Fulton2017, VanEylen2018}. However, alternative hypotheses characterize the demographics of close-in, low-mass exoplanets as a result of gas-accretion in a hot inner disc \citep[e.g,][]{Lee2021, Lee2022}, or separate populations of terrestrial planets and water worlds \citep[e.g,][]{Zeng2019, Venturini2020, Luque2022, Izidoro2022}. 

Direct observations of ongoing atmospheric escape provide an opportunity to test the hydrodynamic escape models proposed to explain the demographics of close-in, low-mass exoplanets. The two most commonly used tracers are Lyman-$\alpha$ \citep[e.g.,][$\sim 10$ detections]{Vidal-Madjar2003, LecavelierDesEtangs2010, Eherenreich2015, Bourrier2017, dosSantos2020} and HeI 10830 \AA~\citep[e.g.,][$\sim 20$ detections]{Spake2018, Allart2018, Kirk2020, ZhangM2023}. Other tracers include H-$\alpha$ or metal lines \citep[e.g.,][]{Vidal-Madjar2004, Sing2019, Ben-Jaffel2022}, however, only a limited number of planets are observed to be undergoing escape in these lines, and therefore they do not probe the broad population of evaporating planets.  

The helium 10830 \AA~line provides an effective avenue to probe escape, however a particular UV stellar spectrum is required to populate the metastable triplet state \citep{Oklopvcic2019}, therefore only a subset of stars (primarily K-stars) are amenable to helium escape observations. In addition, both the fraction of helium in the metastable state (which depends on the typically-unmeasured UV spectrum) and the overall H/He ratio of the escaping gas is poorly constrained. As a result, it is challenging to deduce the outflow properties from observations. 

Lyman-$\alpha$ transit observations are complicated to interpret. Interstellar neutral hydrogen obscures the stellar Lyman-$\alpha$ line core \citep[$\sim -50 \text{ to } 50 \text{ km s}^{-1}$, unless the star has a particularly large radial velocity, e.g.,][]{Landsman1993}, such that to interpret these observations we rely on information contained in the wings of the spectrum. Typical escape models predict outflow velocities to be much smaller than this: $\sim 7-20 \text{ km s}^{-1}$ for photoevaporation and $\sim 1-2 \text{ km s}^{-1}$ for core powered mass loss. Therefore, these observations do not probe the launching regions of the outflow. The fact that transits are indeed observed at such high velocities (generally in the blue wing) indicates that interactions with the circumstellar environment have an important role in making the planetary outflow observable. Furthermore, observations often have a transit so deep it implies a size larger than the planet's Hill sphere and a duration much greater than the optical transit, indicating that a significant absorption comes from parts of the outflow far away (and unbound) from the planet. At these distances, interactions with the circumstellar environment are important \citep[e.g.,][]{Bourrier2013b, Bisalko2013, Matsakos2015, McCann2019}. 

Three mechanisms have been proposed to create high-velocity neutral hydrogen atoms in the outflow: 1) ram-pressure from the stellar wind can accelerate the outflowing gas away from the star; 2) Lyman-$\alpha$ radiation pressure can do the same and 3) charge exchange between stellar wind protons and neutral hydrogen in the outflow can create a population of hot, fast neutral atoms known as energetic neutral atoms (ENAs) \citep[e.g.,][]{Holmstrom2008, Tremblin2012}. There have been a number of 3D particle \citep[e.g.,][]{Bourrier2013b, Bourrier2015, Bourrier2016}, and hydrodynamic/radiation-hydrodynamic \citep[e.g.,][]{Bisalko2013, Matsakos2015, Carroll-Nellenback2016, Khodachenko2019, McCann2019, Debrecht2020, Carolan2021a, Hazra2022, Macleod2022} simulations focusing on the interaction between the planetary outflow and the circumstellar environment. These works show that the planetary outflow is shaped into a cometary tail by a combination of stellar wind ram pressure, radiation pressure and stellar gravity. Typically, these simulations are able to qualitatively reproduce observational features such as the large blue-shifted transits and extended post-transit absorption, indicating that they are capturing the basic physics driving these observations.

Although multi-dimensional hydrodynamics simulations, including radiative transfer and charge exchange, most accurately model Lyman-$\alpha$ transits, direct comparisons between these models and observations have limitations. They are computationally expensive, so they can only be tested over a limited parameter range. They are physically complex, so interpreting how the many uncertain planetary and stellar parameters interplay to produce the resulting transit is challenging. This means that it is unclear whether the inability to match certain observations arises from evaluating models in the wrong part of parameter space or a fundamental issue with physics in the model. 

Motivated by this problem, \cite{Owen2023} developed an analytic model of the cometary tail that predicts how Lyman-$\alpha$ transit observations vary over a large range of possible planetary and stellar parameters. This model highlighted the counter-intuitive nature of Lyman-$\alpha$ observations: stronger EUV irradiation leads to larger mass-loss rates but often weaker transits both in-depth and duration, something previously noted in simulations \citep[e.g.,][]{Bourrier2013, Bourrier2016, Shaikhislamov2020, Villarreal2021}. This provides a simple explanation of why several highly irradiated sub-Neptunes do not have a detectable transit: HD 97658b \citep{Bourrier2017}, $\pi$ Men c \citep{GarciaMunoz2020}, K2-25b \citep{Rockcliffe_2021} and HD 63433b \citep{Zhang2022}.  

Although this model gave a useful insight into how to interpret Lyman-$\alpha$ transits, it simplified the physics of the problem to the extent that it cannot be used to extract quantitative constraints from observations. One problem is that it assumed that the tail remains on the same orbital separation of the planet, which is not the case for very close-in planets like GJ 436 b \citep[e.g.][]{Khodachenko2019}. Furthermore, it assumed that the density of the gas in the tail is constant; however, by conservation of mass, the density should decrease as the gas is accelerated. Finally, the geometry of the tail (i.e. the vertical and horizontal extent of the gas) was calculated by ignoring the stellar wind, which compresses the tail. 

In this work, we build a self-consistent and computationally quick model of the outflowing tail that attempts to accurately reproduce the 3D hydrodynamics simulations that it is based on. We prove the validity of this approach by comparing the trajectory of the tails our model produces to multiple 3D hydrodynamics simulations. We couple this model to a ray tracing routine to produce synthetic transits. The primary motivation of this work is to create a framework to perform ``retrievals'' of atmospheric escape properties on velocity-resolved Lyman-$\alpha$ transit data. This will allow us to elucidate the key physics governing observations, allowing targeted 3D simulations. We emphasize that we do not suggest our model replaces large-scale 3D hydrodynamics simulations. Rather, the model allows us to widely evaluate the parameter space to test whether the physical picture of Lyman-$\alpha$ transits developed in simulations is capable of explaining Lyman-$\alpha$ transits observations and determining the constraining power of these observations. To demonstrate this, we present a study of the GJ 436 b and discuss interpretations of the results. 

\section{Outflow Model}
\label{sec:Tail Model}

In this section, we build a model of the outflowing planetary gas in a steady state. We split the outflow into two components: inside and outside the Hill sphere of the planet. Inside the Hill sphere, the outflow is approximately spherically symmetric \citep[e.g.][]{Murray-Clay2009}; thus, we model the flow as a 1D Parker wind \citep{Parker1958}. Outside the Hill sphere, 3D hydrodynamic simulations show that planetary outflow is sculpted into a cometary tail by the stellar tidal field and stellar wind ram pressure \citep[e.g.,][]{Matsakos2015, Carroll-Nellenback2016, Khodachenko2019, McCann2019, Debrecht2020, Carolan2021a, Hazra2022, Macleod2022}. Inspired by methods used to study disc winds (semi-) analytically \citep{Fukue1990, Clarke2016}, we model the cometary tail as a streambundle, solving for its dynamics. The transition from a spherically symmetric outflow to a cometary tail cannot be modelled semi-analytically. However, the primary goal of this work is to understand the flow far outside the Hill sphere of the planet where a cometary tail is a good approximation. This is because Lyman-$\alpha$ transits primarily probe material outside the planet's Hill sphere. Therefore, we choose to construct the full planetary outflow by ``glueing'' these two outflow models to each other at the Hill sphere of the planet, which we describe in Section~\ref{sec:full_model}. While the transiting Hill sphere region often has a small contribution to the transit profile, this approach allows us to have the outflow into the tail come from a physically motivated (although parameterised) outflow model.    

This model excludes parts of the outflow leading the planet (outside the Hill sphere); therefore, we are not able to model the transit before the ingress of the Hill sphere. We note if there is significant absorption in Lyman-$\alpha$ pre the ingress of the Hill sphere, then our calculated transit depth will be underestimated at ingress and extending through mid-transit, whilst this part of the outflow transits the star. 

For moderate stellar winds, the pre-transit absorption may be highly variable because the gas leading the planet is unstable \citep[e.g.,][]{McCann2019}, which is challenging to model in a semi-analytic framework. As our models are not accurate in this region, we do not compare this part of the transit to observations. We note that for strong incident stellar winds, the pre-transit absorption is not variable \citep[e.g.,][]{McCann2019, Khodachenko2019, Villarreal2021}. Instead, the time of ingress depends on the position of the bow shock, which constrains the ratio between the stellar wind ram pressure and planetary wind ram pressure.  

\subsection{Model of Cometary Tail}
\label{sec:model}

In a frame co-rotating with the planet, we model the cometary tail as a streamline bundle \citep[e.g.,][]{Clarke2016} with an elliptical cross-section. This is composed of a curve representing the trajectory of the tail, with ellipses stacked along it of variable height and depth. The variable ($s$) denotes the distance along the curve. The depth of the ellipse is small compared to the radius of curvature of the tail, and therefore, the tail does not intersect itself. At each point along this curve, we define a left-handed, local Cartesian coordinate system $(l, a, \zeta)$ with the origin at the ellipse's centre. We orient this coordinate system such that $\boldsymbol{\hat{l}}, \boldsymbol{\hat{a}}, \boldsymbol{\hat{\zeta}}$ point tangent to the trajectory of the tail, normal to the trajectory of the tail and parallel to the orbital angular momentum of the planet, respectively. As the tail does not intersect with itself, any point can be specified by a distance along the tail $s$, and coordinates $(a, \zeta)$ in the corresponding ellipse.   

\begin{figure*}
\centering
\includegraphics[width = \textwidth]{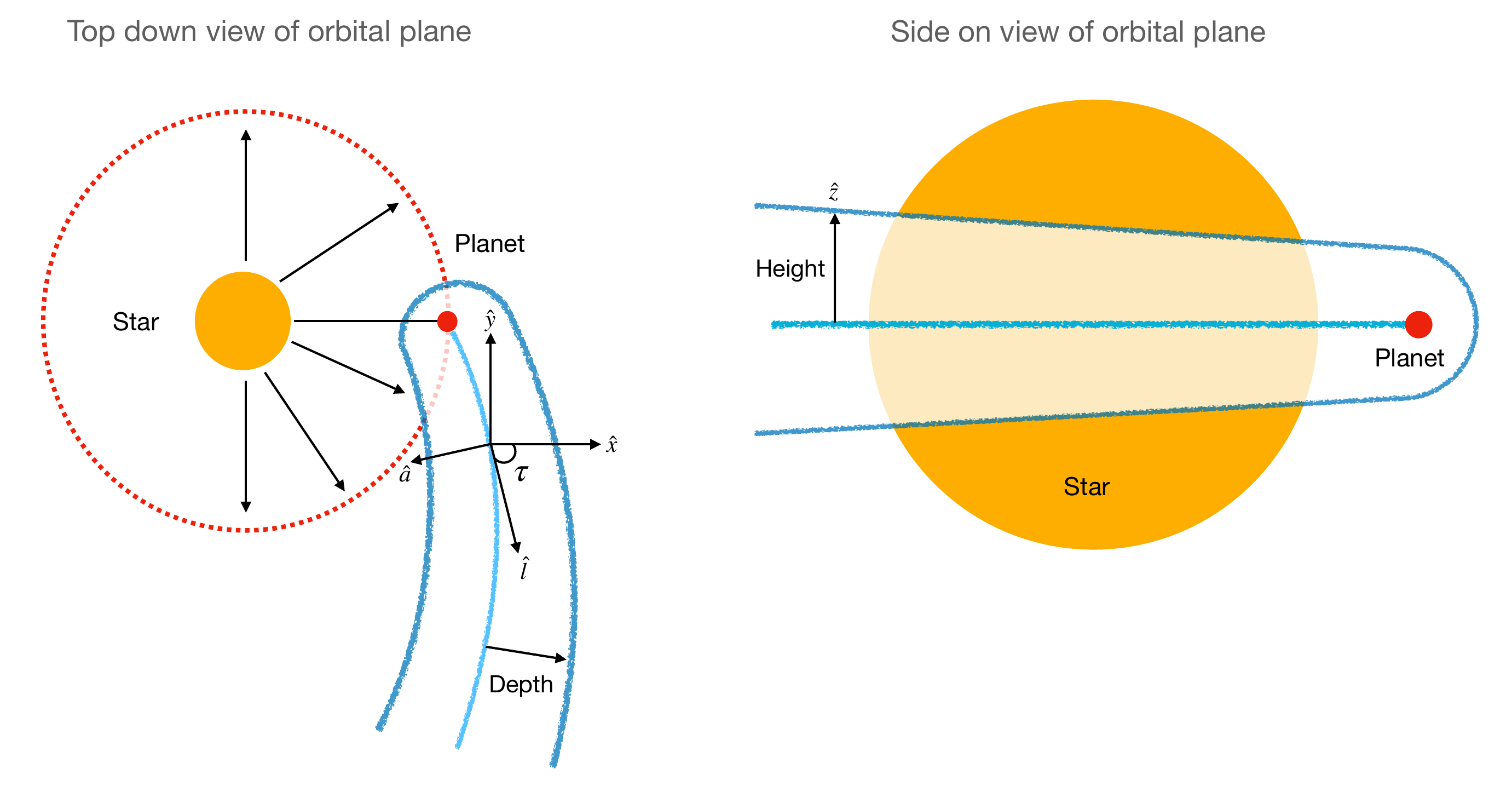}
\caption{A schematic diagram of the planetary outflow that is sculpted into a cometary tail by the stellar tidal field and stellar wind ram pressure. The light blue curve represents the trajectory of the tail, which is parameterised by a coordinate ($s$), representing the distance along this curve. At each point along this curve, we define a local (left-handed) cartesian coordinate system ($l$, $a$, $\zeta$) that can be related to a global cartesian coordinate system ($x$, $y$, $z$) centred on the star. We solve for i) the trajectory, ii) the ionization state, and iii) the geometry of the tail as a function of system parameters shown in Table~\ref{tab:Fixed and Free Parameters}.}
\label{fig:tail_schematic}
\end{figure*}

We also define a global right-handed Cartesian coordinate system $(x, y, z)$ co-rotating with the planet, with its origin at the star. The radial distance from the star is denoted by $r$. We choose the $(x, y)$ coordinates to lie in the planet's orbital plane so that the trajectory of the tail also lies in the orbital plane (i.e. $s = s(x, y)$). Given a point along the tail $s_0 = (x_0, y_0)$, a point in the corresponding cross-sectional plane is given by:   

\begin{align}
    &x = x_0 - a\text{sin}\tau\\
    &y = y_0 - a\text{cos}\tau\\
    &z = \zeta
\end{align}

\noindent where $\tau$ is the angle between $\boldsymbol{\hat{l}}$ and $\hat{\textbf{x}}$ measured anticlockwise. Figure \ref{fig:tail_schematic} shows a sketch of the model and coordinate systems. For simplicity, we consider both the velocity and the ionization fraction of the gas in the tail to be a function of the distance $s$ along the tail only. These simplifications allow us to separately model the bulk motion (i.e. the trajectory), ionisation state, geometry and density structure of the outflow. The model, therefore, consists of three parts:\\

\noindent \textbf{(i)} Self-consistently solving for the trajectory of the tail in the planet's orbital plane by assuming a generalised geostrophic balance.\\  

\noindent \textbf{(ii)} Solving for the neutral fraction of hydrogen atoms along the tail using the optically thin ionisation equation.\\  

\noindent \textbf{(iii)} Solving for the height, depth and density structure of this streambundle, assuming the gas is in hydrostatic equilibrium in the cross-section of the streambundle.\\      

Modelling the tail as a streambundle implicitly treats the planetary outflow as a fluid as opposed to the particle approach used in some simulations of the outflow \citep[e.g.,][]{Bourrier2013b, Eherenreich2015, Bourrier2015}. At $10^4$ K, the proton-hydrogen cross section is $\sim 10^{-14} \text{ cm}^{2}$ \citep{Schultz2008}. Typically, proton number densities in the planetary outflow are $\gtrsim 10^5 \text{ cm}^{-3}$ \citep[e.g.][]{McCann2019}, therefore the mean free path for a neutral hydrogen is $\lesssim 10^9 \text{cm}$, which is generally small compared to the scale of the outflow, therefore treating the gas as collisional is valid. We check this condition in our simulated outflows post-facto to ensure that it holds. 

\subsubsection{Trajectory of the Tail}

In a reference frame co-rotating with the planet, in which we treat the flow as steady, we solve for the trajectory of the tail as a 1D streamline subject to momentum conservation.    

\begin{align}
    (\textbf{u} \cdot \boldsymbol{\nabla})\textbf{u} = \textbf{a}_{\text{sw}} - \boldsymbol{\nabla}\phi - 2\boldsymbol{\Omega} \times \textbf{u}
    \label{momentum_conservation_eq}
\end{align}
where $\textbf{a}_{\text{sw}}$ is the acceleration due to ram pressure of the stellar wind, $\boldsymbol{\nabla}\phi$ is effective potential, and $2\boldsymbol{\Omega} \times \textbf{u}$ is the Coriolis force. We assume the planet has a circular orbit and ignore the Euler force (it's not possible to construct a steady-state model of an eccentric orbit). Although the stellar wind ram pressure force is a surface force, we approximate that this force is applied like a body force, as pressure gradients quickly distribute this force across the tail. Furthermore, geostrophic balance keeps the tail stable. We discuss the validity of this approximation later in this section. In this framework, the force per unit length that the stellar wind applies to the tail is:

\begin{align}
&\frac{d\textbf{p}}{dt} = 2H\rho_{*}(\textbf{u}_{*} - \textbf{u}\,\text{cos}\chi)^2\text{sin}\chi \: \hat{\textbf{u}}_{*} 
\end{align}
where $H$ is the height of the tail, $\rho_{*}$ and $\textbf{u}_{*}$ are the stellar wind density and velocity, respectively, and $\chi$ is the angle between the tangent of the tail and the direction stellar wind. In the case where the stellar wind is strong, simulations show that all the mass lost from the planet is directed down the cometary tail \citep[e.g.,][]{McCann2019}. Therefore, we can define the mass per unit length in the tail:

\begin{align}
\lambda = \frac{\dot{M}_p}{u}
\end{align}
where $\dot{M}_p$ is the mass loss rate of the planet and $u = |\textbf{u}|$. Therefore, the acceleration of the tail due to stellar wind ram pressure is given by:   

\begin{align}
\label{eq:tail_acceleration}
\textbf{a}_{\text{sw}} = \frac{2Hu\rho_{*}(\textbf{u}_{*} - \textbf{u}\,\text{cos}\chi)^2\text{sin}\chi}{\dot{M}_p} \hat{\textbf{u}}_{*}  
\end{align} 
For this framework to be valid, the timescale for pressure waves to distribute this force across the tail must be shorter than the timescale in which the forces acting on the tail significantly change. We estimate this by comparing the  acceleration timescale to the sound crossing time as:

\begin{align}
\frac{\tau_{sc}}{\tau_{a}} = \frac{\text{dlog}u}{\text{dt}}\frac{2D}{c_s}
\end{align}
where $c_s$ is the isothermal sound speed of the gas in the outflow and $D$ is the depth of the tail. Taking the conservative assumption that $u_{*} \gg u$ and $\text{sin}\chi = 1$, so that $a_{\text{sw}} \sim (2Hu_{*}^2\rho_{*})/\dot{M}_p$, this equals 

\begin{align}
\frac{\tau_{sc}}{\tau_{a}} \lesssim \frac{1}{\pi}\left(\frac{\dot{M}_*}{\dot{M}_p}\right)\left(\frac{u_*}{c_s}\right)\left(\frac{HD}{r^2}\right)
\end{align}

\noindent For a typical sub-Neptune around a solar-like star at 0.03 AU, this is smaller than one.

\begin{align}
\begin{split}
\frac{\tau_{sc}}{\tau_{a}} \lesssim \ & \left(\frac{\dot{M}_*}{10^{12} \text{ g s}^{-1}}\right)\left(\frac{\dot{M}_p}{10^{10} \text{ g s}^{-1}}\right)^{-1}\left(\frac{u_*}{100 \text{ km s}^{-1}}\right)\\
& \times \left(\frac{c_s}{10 \text{ km s}^{-1}}\right)^{-1}\left(\frac{HD}{3 \times 10^{-3}r^2}\right)
\end{split}
\end{align}
where $H$ and $D$ are conservatively estimated using the distance that a ballistic particle travels after being launched from the planet at 10 km s$^{-1}$.     

As this estimate is an upper limit, it is reasonable to approximate the total effect of stellar wind ram pressure to apply a force equal to the momentum flux throughout the tail. Additionally, in this framework, the height and depth scales like the bow shock radius. Therefore, increasing the stellar wind mass loss rate or velocity should reduce the height and depth equivalently so that the argument still holds. 

We also test the validity of this approach by comparing the tail trajectories that we have calculated in our model to multiple 3D hydrodynamic simulations (see Section~\ref{sec:Comparison to Hydro} and Figure~\ref{fig:Comparison_to_hydro}). Detailed inspection of publicly available simulation outputs demonstrates that we are able to replicate the trajectories of these simulations using the same system parameters, indicating that our assumptions are reasonable.    

\subsubsection{Geometry of the Tail}

A corollary of the previous argument is that the gas inside the tail should adjust to its local conditions. Therefore, in the elliptical cross-section of the tail, we can approximate the gas to be in hydrostatic equilibrium. Perpendicular to $\boldsymbol{\hat{l}}$, the forces on a parcel of gas are:  

\begin{align}
 -\frac{1}{\rho(a,z)}\boldsymbol{\nabla}{P(a,z)} - \boldsymbol{\nabla}\phi(a,z) - 2\boldsymbol{\Omega} \times \textbf{u} + \frac{\textbf{u}^2}{r_c(a)}\hat{\textbf{a}} = 0
\end{align}
where $\rho$ and $P$ are the density and pressure, respectively, and $r_c$ is the radius of curvature corresponding to streamlines in the outflow. We note that this equation is also an implicit function of the distance along the tail $s = (x, y)$. The effective potential is: 

\begin{align}
\label{eq:effective_potential}
\begin{split}
\phi = -&\frac{GM_{*}}{\sqrt{(x - a\text{sin}\tau)^2 + (y - a\text{cos}\tau)^2 + z^2}}\\ 
&- \frac{1}{2}\Omega^2\left((x - a\text{sin}\tau)^2 + (y - a\text{cos}\tau)^2\right)
\end{split}
\end{align}
Since the height and depth of the tail are small compared to the distance from the star, we expand the effective potential and radius of curvature to second-order in terms of $\left(\frac{a}{r}, \frac{z}{r}\right)$. Up to this order, the horizontal ($a$) and vertical ($z$) coordinates decouple; therefore, the density is separable, and each direction can be solved independently. For simplicity, we solve for the density using an isothermal equation of state such that $P = c_s^2 \rho$. In the vertical direction, the density distribution is given by     

 \begin{align}
 -\frac{c_s^2}{\rho(z)}\frac{\partial \rho(z)}{\partial z} - \frac{GMz}{r^3} = 0
\end{align}
which can be analytically solved as 
\begin{align}
\rho(z) = \rho_{a,0}\text{exp}\left[-\frac{GMz^2}{2c_s^2r^3}\right] = \rho_{a,0}\text{exp}\left[-\frac{\Omega^2_kz^2}{2c_s^2}\right]
\end{align}
 where $\rho_{a,0}$ is the density in the orbital plane at horizontal coordinate $a$ and $\Omega_k$ is the Keplerian angular velocity. This is just the standard solution for vertical hydrostatic equilibrium of a rotating fluid (e.g. in a disc).      

Following the same procedure to calculate the horizontal density distribution leads to unphysical solutions. This is a result of our assumption that the velocity of the gas is constant across the cross-sectional slice of the tail. Instead, we approximate the depth by assuming that the angular momentum difference between the edge of the tail and the centre stays constant, where the initial dispersion is approximated by launching particles at the sound speed with varying angular momenta from the planet. In the limit where the orbital velocity of the planet is greater than the velocity of the tail, the depth $\sim \frac{c_s}{\Omega_k}\sqrt{\frac{a_p}{r}} \sim \frac{c_s}{\Omega_k}$. This agrees with the typical depths found in simulations for weak incident stellar winds. For stronger stellar winds, the depth is smaller as the stellar wind compresses the tail. Therefore, a sensible choice for a density distribution is: 

\begin{align}
\rho(a) \propto \text{exp}\left[-\frac{\Omega_k^2a^2}{c_s^2}\right]
\end{align}
{We note that in order to calculate the optical depth of the tail in transit, we integrate this density profile (almost) horizontally through the tail, and therefore, the transit depth will only be weakly dependent on this choice.} Given this choice, the total density structure is given by:    

\begin{align}
\label{eq:total_density}
    \rho = \rho_0\text{exp}\left[-\frac{a^2}{\alpha^2} - \frac{z^2}{\beta^2}\right], \quad \alpha = \frac{c_s}{\Omega_k}, \quad \beta = \frac{\sqrt{2}c_s}{\Omega_k} 
\end{align}
where $\rho_0$ is the density at the centre of the ellipse. In order to find the height, depth and $\rho_0$ of the tail, three constraints must be applied. The first of these comes from equating the mass-flux to the integral of the density over the elliptical cross-section. The latter two are given by specifying conditions at the boundary between the planetary and stellar wind. As the stellar wind is incident on the inner edge of the tail, conditions vary substantially on either side of the tail. We choose to evaluate this at the inner edge and top of the ellipse (we could equivalently define it a the bottom), where these boundaries are well-defined. These three constraints can be written as:

\begin{align}
&\text{i)} \ \iint_{\mathcal{A}} \rho(a,z)\,dadz = \frac{\dot{M}_p}{u} \quad \text{ (Mass per unit length)} \label{eq:density_mass_conservation}\\
&\text{ii)} \ \ \rho(-D, 0)c_s^2 = P_{\text{sw},D} \quad \text{ (Pressure at inner edge)}\label{eq:depth_boundary}\\ 
&\text{iii)} \ \rho(0, H)c_s^2 = P_{\text{sw}, H} \quad \text{ (Pressure at upper edge)}
\label{eq:height_boundary}
\end{align} 
where $\mathcal{A}$ is the area of the ellipse and $P_{\text{sw},D}$ and $P_{\text{sw}, H}$ is the stellar wind pressure at the inner edge and top of the tail respectively. We note that the obtained stellar wind pressure should correspond to the shocked stellar wind. In the orbital plane, we approximate the shock front to be parallel to the trajectory of the tail, and in the vertical plane, it forms a bow shock. Since simulations indicate that the bow shock is not thin, there are no general analytic ways to calculate the shape of a bow shock or the downstream pressure. However, it is possible to solve for the pressure at the nose of the shock using the oblique shock equations. For simplicity, we assume that the pressure at the inner edge and top of the tail are equal and are a linear scaling of the pressure at the nose. 

\begin{align}
\kappa P_{\text{nose}} = P_{\text{sw},D} = P_{\text{sw}, H}
\end{align}

As we expect the pressure to decrease as the gas moves away from the nose, $\kappa < 1$. We set $\kappa = 0.3$ using \citet{Schulreich2011} as a guide. With this simplification, using Equations~\ref{eq:total_density}, \ref{eq:depth_boundary}, \ref{eq:height_boundary}, the depth and height are related as: 

\begin{align}
\label{eq:hd_relation}
\frac{D}{\alpha} = \frac{H}{\beta}
\end{align}
The integral in Equation~\ref{eq:density_mass_conservation} cannot generally be solved analytically. However, it can be written in terms of special functions \citep[e.g][]{Waugh1961}.  

\begin{align}
\frac{1}{2\pi}\iint_\mathcal{A} e^{-\frac{x^2 + y^2}{2}} dxdy = p\left(\frac{A + B}{2}, \frac{A - B}{2}\right) - p\left(\frac{A - B}{2}, \frac{A + B}{2}\right)
\end{align}

\begin{align}
p(R, r) = e^{-\frac{r^2}{2}}\int_{0}^{R}e^{-\frac{t^2}{2}}I_0(rt)tdt
\end{align}
where $A$, $B$ are the semi-major and semi-minor axes of the ellipse, $\mathcal{A}$ is the area of the ellipse and $I_0$ is the modified Bessel function of the first kind. In the special case given by Equation~\ref{eq:hd_relation}, this can be solved exactly to give:  

\begin{align}
&D = \alpha\text{ln}\left(\frac{\dot{M}_pc_s^2}{u\pi \gamma P_{\text{nose}}\alpha\beta} + 1\right)\\
&H = \frac{\beta}{\alpha}D
\end{align}

\subsubsection{Ionization of Tail}

As absorption of Lyman-$\alpha$ radiation is from neutral hydrogen, it is essential to calculate the ionization state of hydrogen in the tail. In the photoevaporation model, where the absorption of EUV photons launches the planetary outflow, the tail must be optically thin to EUV photons; otherwise, the EUV flux absorbed at the base of the flow would be too small to drive a large outflow \citep{Owen2023}. \citet{Owen2023} showed that this extends to core-powered mass loss outflows, which are not driven by EUV radiation. As a result, the photoionization rate of neutral hydrogen is approximately constant in each cross-sectional slice of the tail (also noting that $D, H \ll r$).

We also include hydrogen recombination in our model. For old small planets, whilst observable, the gas in the tail is far way from photoionization-recombination equilibrium, and therefore, the effect of recombination is minimal. However, for giant planets or young planets with high mass loss rates, the outflows may be dense enough for recombination to be important. As recombination is density-dependent, it varies down the tail. For simplicity, we approximate that the recombination rate of neutral hydrogen is constant in each cross-sectional slice of the tail and equivalent to the mean density in that slice. The number density, $n$, of protons and hydrogen atoms in a cross-sectional slice of the tail is:

\begin{align}
n = \frac{\dot{M}_p}{\pi uHDm_H}
\end{align}
where $m_H$ is the mass of a hydrogen atom. As such, we can simplify the ionization fraction to a 1D problem, which depends on distance $(s)$ down the tail. In this limit, the ionization equation as a parcel of gas moves down the tail can be written as:

\begin{align}
    \frac{DX}{Dt} = (1 - X)\Gamma - nX^2\alpha_{A}
    \label{ionisation_eq}
\end{align}

\noindent where X is the ionisation fraction, $\Gamma$ is the photoionisation rate and $\alpha_{A}$ is the case A recombination coefficient. We use the case A recombination coefficient because the tail is optically thin to recombination photons. This is because photons with energies $> 13.6 \text{ eV}$, that are produced by a hydrogen atom recombining directly to the ground state generally escape the tail rather than ionizing another hydrogen atom. 

\noindent The optically thin photoionization rate is given by:

\begin{align}
    \Gamma = \int_{13.6~{\rm eV}}^{\infty}\frac{L_{\nu}}{4\pi r^2h\nu}\sigma_{\nu}d\nu
\end{align}

\noindent where $\nu$ is the frequency of a photon, $\sigma_{\nu}$ is the photoionization cross section of hydrogen as a function of frequency, and $L_{\nu}$ is the luminosity of the star as a function of frequency. To map the optically thin photoionization rate to an EUV luminosity, one needs to use a stellar spectrum (see Appendix~\ref{appendix2}). The optically thin photoionization rate is the physical quantity that determines the ionization structure in the tail, we specify this quantity in our model rather than a stellar EUV luminosity. 

Thus, in steady state, the governing equation for the ionization fraction becomes:

\begin{align}
\frac{dN}{ds} = -\frac{\Gamma N}{u} + \frac{n(1-N)^2\alpha_{A}}{u}
\end{align}
where $N = 1 - X$ is the neutral fraction of hydrogen. 

\subsubsection{Stellar Wind}

In order to calculate the acceleration of the tail due to the stellar wind, we must specify the velocity and density of the stellar wind. For simplicity, we treat the stellar wind as spherically symmetric, such that we can link the density, velocity, and mass loss rate of the star through mass conservation.

\begin{align}
\dot{M}_{*} = 4\pi r^2\rho_*u_*
\end{align}
where $\dot{M}_{*}$ is the mass loss rate of the star. In this exploratory work, we also choose the stellar wind to impact the tail radially. We note that the outflow model, along with the simulations it is based on, is only valid for planets located outside the Alfvén surface of the star where the ram pressure of wind dominates over the magnetic pressure of the star. The velocity structure of the stellar wind can be specified as an input parameter, which can be taken from stellar wind simulations. For simplicity, in this paper, we will take the stellar wind velocity to be constant as the tail gets radially pushed outwards. This is valid if the planet lies outside the acceleration zone of the stellar wind. We note that many evaporating planets are very close to their stars, such that the stellar wind is still accelerating. 

With this approximation, the ram pressure of the stellar wind is:    

\begin{align}
\rho_{*}u_{*}^2 = \frac{\dot{M}_{*}u_{*}}{4\pi r^2}     
\end{align}
Although we have ignored many of the intricacies of the stellar wind, we highlight that Lyman-$\alpha$ transits are probing the stellar wind conditions near the planet. A possible complication with this arises when stellar wind conditions can vary considerably along the orbit of the planet. However, more complicated stellar wind conditions could be added to the model relatively simply. 

\subsubsection{Method of Solution}

It appears that we have to solve a system of coupled partial differential equations for the trajectory, ionization fraction, and geometry of the tail. However, as the domain of all the functions in these equations is the 1D trajectory curve, they can be written as a function of distance along the tail $(s)$. This system can equivalently be written as a set of coupled ordinary differential equations:     

\begin{align}
    &u\frac{du_x}{ds} = (\textbf{a}_{\text{sw}})_{x} - (\boldsymbol{\nabla}\phi)_{x} + 2\Omega u_y\\
    &u\frac{du_y}{ds} = (\textbf{a}_{\text{sw}})_{y} - (\boldsymbol{\nabla}\phi)_{y} - 2\Omega u_x\\
    &\frac{dx}{ds} = \frac{u_x}{u}\\
    &\frac{dy}{ds} = \frac{u_y}{u}\\
    &\frac{dN}{ds} = -\frac{\Gamma N}{u} + \frac{n(1-N)^2\alpha_{A}}{u}\\
    &D = \alpha\text{ln}\left(\frac{\dot{M}_pc_s^2}{u\pi P_{\text{sw},D}\alpha\beta} + 1\right)\\
    &H = \frac{\beta}{\alpha}D
\end{align} 

\noindent where the notation $(\textbf{A})_{i}$ represents the component of the vector $\textbf{A}$ in the $i$ direction. We solve these implicitly using an implementation of a backward differentiation formula method BDF \citep{Byrne1975} provided in the \textit{scipy} library \citep{scipy}. The relative and absolute tolerance used were $1 \times 10^{-13}$ and $1 \times 10^{-14}$, respectively. 

\subsection{Spherical Wind}

Inside the Hill sphere, the outflow is modelled as a 1D constant sound speed Parker wind, with the addition of tidal gravity, along a streamline originating from the sub-stellar point on the planet. We do this in order to stay agnostic about the mechanism driving the outflow. The outflow is parameterised by a mass loss rate and a sound speed that is constant throughout the flow. For a given mass loss rate and sound speed, the velocity of the outflowing gas is analytically given by:

\begin{align}
\label{eq:velocity}
&u = 
\begin{cases}
-W_0[-D(r')] \quad \ r' \leq r_{cr}\\
-W_{-1}[-D(r')] \quad r' > r_{cr}
\end{cases}
\end{align}

\begin{align}
D(r') = \left(\frac{r'}{r_{cr}}\right)^{-4}\text{exp}\left[4r_{\alpha}\left(\frac{1}{r_{cr}} - \frac{1}{r'}\right) + \frac{2r_{\alpha}}{R_H^3}(r_{cr}^2 - r'^{2}) - 1\right]
\end{align}

\noindent where $r'$ is the radial distance from the planet, $W_k$ is the Lambert W function on branch $k$ \citep{Cranmer2004}, $r_{cr}$ is the sonic point of the flow, $r_{\alpha}$ is the sonic point for the flow without tidal gravity and $R_H$ is the Hill radius of the planet. We calculate the fraction of ionized hydrogen by analogy with our method for the ionizing fraction along the tail. Thus we use:

\begin{align}
\label{eq:spherical_ionisation_eq}
\frac{\partial N}{\partial r'} = -\frac{\Gamma e^{-\tau_{\bar{\nu}}(r')} N}{u} + \frac{\dot{M}_p(1-N)^2\alpha_{A}}{4\pi r'^2u^2m_p}
\end{align}
We include the optical depth of the outflow to $20$ eV photons, $\tau_{\bar{\nu}}$, in the photoionization part of the equation to account for the fact that the outflow can be optically thick near the planet. Using a single photon energy, rather than a stellar spectrum, to calculate the optical depth does not give an accurate ionization fraction deep in the atmosphere of the planet where the wind is launched \citep[e.g.,][]{Trammell2011}. We note that the overall transit depth is dominated by the flow at larger radius, where the outflow is optically thin to EUV radiation. This is because the outflow is still generally optically thick to Lyman-$\alpha$ photons (even in the observable wings) at these larger radii due to large difference between the EUV and Lyman-$\alpha$ cross sections ($\sim$ 4 orders of magnitude), despite the decreasing density of neutral hydrogen. As the optically thin region (to EUV radiation) covers a significantly larger area of the star than the optically thick region, it dominates the contribution to the transit depth. Therefore, we are primarily concerned with approximating the ionization fraction in the optically thin region well. As we specify the optically thin ionization rate in our model, we approximate the ionization fraction in this region well.

Equation \ref{eq:spherical_ionisation_eq} has to be solved iteratively as $\tau_{\bar{\nu}}(r')$ depends on the ionization fraction of the outflow. However, we can approximate $\tau_{\bar{\nu}}(r')$ by assuming the atmosphere is neutral and solving the differential equation directly \citep[e.g.,][]{DosSantos2022}. We solve this using an implementation of the LSODA method \citep[][]{Petzold1983} provided in the \textit{scipy} library \citep{scipy}. The absolute and relative tolerances used are $1 \times 10^{-13}$ and $1 \times 10^{-13}$ respectively.

Although modelling the outflow inside the Hill sphere as a Parker wind is not as accurate as a self-consistent EUV heated \citep[e.g.,][]{Murray-Clay2009} or bolometrically heated \citep[e.g.,][]{Schulik2023} model, we emphasize that the primary motivation of modelling the Hill sphere is to find the initial conditions of the gas entering the tail. As mentioned previously, this also allows us to remain agnostic to the exact physics that is driving the outflow. This agnostic nature then allows us to test the prediction of various classes of escape models (e.g., photoevaporation vs core-powered mass-loss).     

\subsection{Energetic Neutral Atoms}

 The interaction between the stellar wind and the planetary outflow also results in charge exchange between the stellar wind protons and neutral hydrogen in the outflow. There is little momentum transfer between the atoms in this process \citep[e.g.,][]{Pinto2008}, and therefore it generates a high-velocity neutral atom. The production of these high-velocity neutral atoms, known as energetic neutral atoms (ENAs), has been proposed as the source of the large Lyman-$\alpha$ transits in some systems \citep[e.g.,][]{Holmstrom2008, Kislyakova2014, Bourrier2016, Khodachenko2019}. 

 There are a range of approaches used to model the generation of ENAs in the literature: particle-based \citep[e.g][]{Holmstrom2008, Bourrier2016}, single-fluid \citep[e.g.][]{Tremblin2012, Debrecht2022} or multi-fluid \citep[e.g][]{Khodachenko2019}. In this work, we adopt a single-fluid description of the production of ENAs. This assumes that stellar wind protons cannot penetrate into the planetary outflow and are forced around the flow. The resulting velocity shear at the boundary between the stellar and planetary wind creates Kelvin-Helmholtz instabilities turbulently mixing the stellar wind and planetary gas. Charge exchange between stellar wind protons and planetary neutral hydrogen atoms in this turbulent mixing layer generates ENAs. 
 
 The single fluid framework is valid when the mean free path of stellar wind protons in the planetary outflow is small compared to the scale of the planetary outflow. To ensure that we are justified in using this framework, we post-facto check that these conditions hold for our simulated outflows, shown in Section \ref{sec:ss}
 
We estimate the number density of ENAs in the turbulent  mixing layer using analytic arguments from \citet{Tremblin2012, Owen2023}. The rate of production of ENAs is given by:   

\begin{align}
\beta(n_{+,*}n_{0, p} - n_{0, *}n_{+, p})
\end{align}
where $n_{+,*}$ and $n_{+,p}$ are protons of stellar wind and planetary wind origin respectively, $n_{0,*}$ and $n_{0,p}$ are neutral hydrogen atoms of stellar and planetary origin respectively and $\beta$ is the charge exchange rate coefficient. In the mixing layer where the stellar and planetary wind are interacting, the collisional timescale is much shorter than the timescale for ENAs to get advected off the limb of the star, we can assume that the stellar wind protons and planetary atoms are in chemical equilibrium. Therefore, the number density of ENAs in the mixing layer can be approximated as: 

\begin{align}
n_{\text{ENA}} = Nn_{*}
\end{align}
where $n_{*}$ is the density of the post-shocked stellar wind and $N$ is the neutral fraction of hydrogen in the tail.  

The contribution of ENAs to the Lyman-$\alpha$ optical depth also depends on the velocity and temperature of these atoms and the size of the mixing layer. The collision and subsequent diversion of the stellar wind around the tail may cause the bulk velocity of ENAs in the mixing layer to be different to the initial stellar wind velocity. As we do not have an a priori way of calculating the bulk velocity of ENAs, we treat it as a constant that is allowed to be a free parameter of the model. Elastic collisions between atoms of stellar and planetary origin in the mixing layer cool the atoms of the stellar origin. For simplicity, we will ignore this process and set the temperature of ENAs to be the same as the stellar wind temperature. 

\citet{Raga1995} analytically studied the size of the mixing layer for the interaction between a spherical outflow with a planet parallel flow. To do this calculation, they had to assume that the stand-off distance between the two shocks was small compared to the radius of curvature of the shock front. This is not the case for the interaction between the stellar wind and the planetary outflow, where the shocked regions can be thick. However, using this as a guide and inspecting the simulations performed by \citet{Tremblin2012, Debrecht2022} allows us to estimate the width of the mixing layer to range from a few to tens of per cent of the depth of the tail. As this range is fairly large, we choose to treat the size of the mixing layer as a free parameter, specified by $L_{\text{mix}}$, which represents the fractional size of the mixing layer compared to the radius of curvature of the planetary outflow that the stellar wind impinges on.       

For ENAs generated due to the interaction of the stellar wind with the Hill sphere of the planet, the mixing layer is given by the area between spheres with Hill sphere radius $r_H$ and $r_H(1 + L_{\text{mix}})$. For ENAs generated in the tail, the mixing layer is given by the area between an ellipse with a height and depth given by $H, D$ and an ellipse with a height and depth given by $H(1 + L_{\text{mix}}), D(1 + L_{\text{mix}})$ respectively.     

\subsection{Constructing the Full Outflow Model}
\label{sec:full_model}

In order to construct the full outflow model, we need a way to transition from the spherically symmetric outflow inside the Hill sphere into the cometary tail-like outflow. We do this by "glueing" the tail onto a point on the Hill sphere of the planet. The initial speed and ionization fraction of the gas in the tail are specified by the values given by the spherical wind model evaluated at the Hill sphere. With this formulation, there is still freedom to choose exactly at what position to glue the tail onto the Hill sphere and the initial direction of the outflow. To simplify this, we assume that the initial position and velocity vector of the tail, in a frame centred on the planet, are parallel. We can then encode this information in a free parameter $h$, which we interpret as the relative specific angular momentum between the outflow and the planet. We shall refer to this quantity as the angular momentum deficit of the outflow, and it can be written as: 

\begin{align}
\label{eq:ang mom def}
h = \Omega(a_p^2 + 2a_pR_H\text{sin}\theta + R_H^2) + u\text{cos}\theta\sqrt{a_p^2 + 2a_pR_H\text{sin}\theta + R_H^2}
\end{align}
where $\theta$ is the angle at which the tail is glued onto the Hill sphere of the planet, and $a_p$ is the semimajor axis of the planet. 
We estimate the projected area of this transition region blocking the star relative to the rest of the tail is $\sim \frac{R_H^2}{R_{*}H}$, which is generally small, and therefore this uncertainty should not have a large effect on our results. Furthermore, the primary aim of this work is to model the transit on scales much larger than the Hill sphere of the planet where the geometry of the model matches 3D simulations. 

The full outflow model requires 14 input parameters, shown in Table~\ref{tab:Model_Parameters}. These can be roughly divided into "tightly constrained" and "weakly constrained" parameters. The "tightly constrained" are ones, like the planet's or the star's radius, that are generally well constrained by independent observations. In general, when comparing to observations, we would allow the "weakly constrained" parameters to be free and the "tightly constrained" parameters to be fixed or use these independent observations as tight priors.     

\begin{table}
  \label{tab:Model_Parameters}
  \caption{The input parameters for our tail model. The fixed parameters are well constrained from other observations. The free parameters are the ones we would like to infer from observations, particularly the mass loss rate and sound speed of the escaping gas. }
  \begin{tabular}{p{3.2cm}p{4.8cm}}
   Tightly Constrained & Weakly Constrained\\
   \hline
   Planet radius, $R_p$ & Sound speed, $c_s$ \\
   Planet mass, $M_p$ & Planet mass loss rate, $\dot{M}_p$ \\
   Stellar mass, $M_{*}$ & Stellar wind velocity (at planet), $u_{*}$\\
   Semimajoraxis of orbit, $a_p$ & Star mass loss rate, $\dot{M}_{*}$\\
   Orbital inclination, $i$ & Optically thin photoionization rate (at planet), $\Gamma_p$\\
   & Stellar wind temperature, $T_{\text{sw}}$\\
   & Angular momentum deficit of outflow, $h$\\
   & Bulk velocity of ENAs, $u_{\text{ENA}}$\\
   & Width of turbulent mixing layer, $L_{\text{mix}}$\\
   \end{tabular}
\end{table}

\section{Comparison to 3D Hydrodynamics Simulations}
\label{sec:Comparison to Hydro}

To demonstrate that this model is accurate in predicting the geometry of the outflowing gas, we compare the trajectory that our model predicts to those obtained from 3D hydrodynamics simulations. We do this for four simulation setups described in \citet{McCann2019, Khodachenko2019, Macleod2022}; and \citet{Hazra2022}. These simulations examine the outflow for a range of planetary and stellar properties to highlight the model's applicability over a wide range of parameters. Specifically, the masses of the planets we compare to range from $0.07-1.14 \ M_J$, with mass loss rates ranging from  $0.2-10 \times 10^{10} \text{ g s}^{-1}$ and stellar type ranging from M to F. 

For the comparison to be meaningful, we fix the free parameters in our model to be the same as those used in the respective simulation. In each case, the planetary mass loss rate, stellar wind mass loss rate, and stellar wind temperature are provided in the simulation paper. They also provide a temperature plot of the outflow, which we can use to estimate the sound speed of the gas. These are estimated to be 11, 11, 9, and 6 km s$^{-1}$ respectively. We fix the value of the stellar wind velocity to match the value of stellar wind velocity at the planet's position in the simulations. Finally, the angular momentum deficit, $h$, is not fixed from the initial simulation parameters, so we approximate it by looking at the direction of the gas flow at the beginning of the tail. Normalized to the specific angular momentum of the planet, the angular momentum deficit is given by -0.05, 0.07, 0.02, and -0.06, respectively.

\begin{figure*}
\centering
\includegraphics[width = 0.8\textwidth]{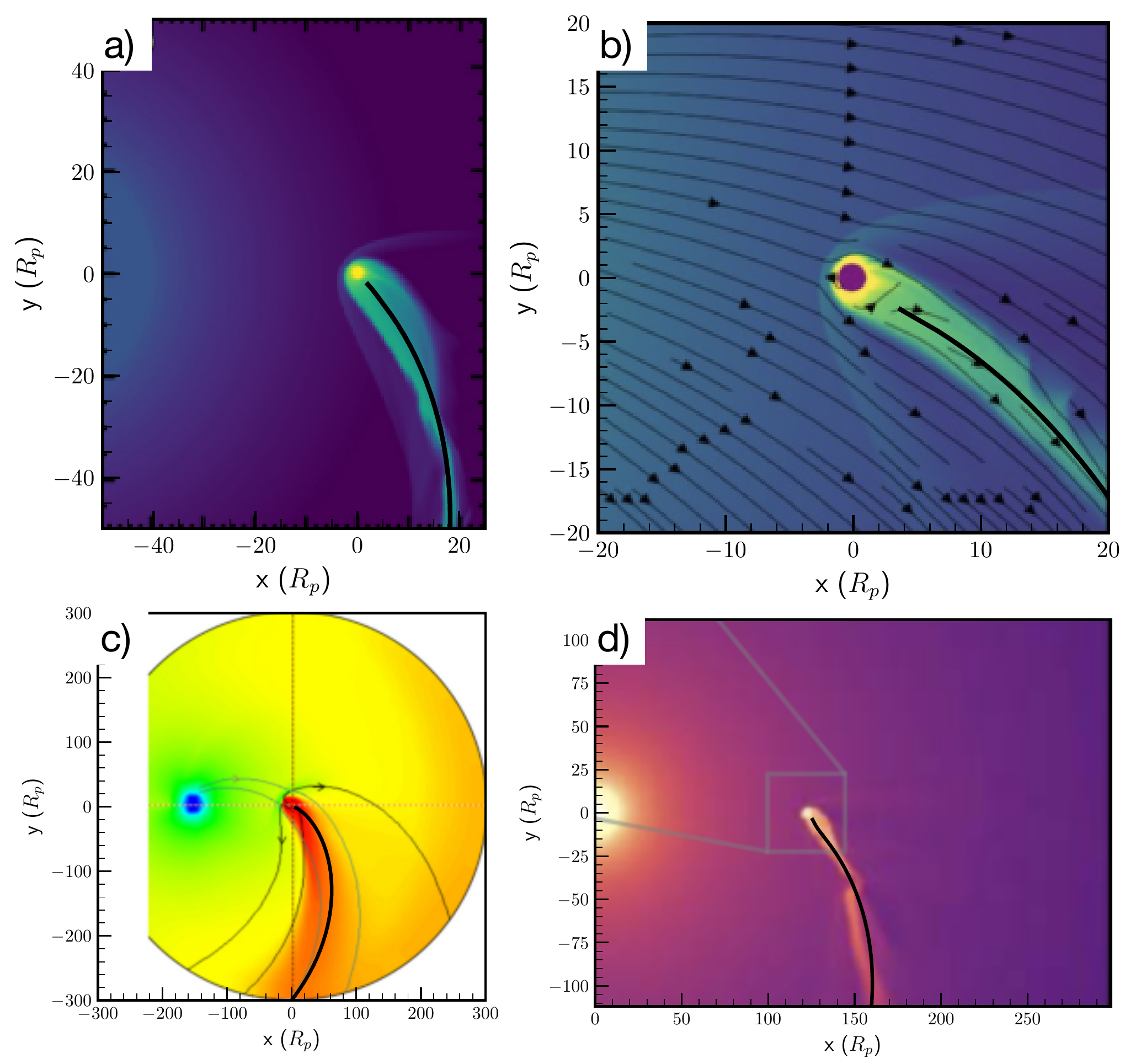}
\caption{Each panel shows the geometry of the planetary outflow produced from 3D hydrodynamics simulations, covering a variety of planet and system parameters. These are: a) Hot Jupiter based on HD209458b \citep{McCann2019}, b) HD189733b, a Hot Jupiter \citep{Hazra2022} , c) GJ436b, a Warm Neptune \citep{Khodachenko2019} and d) Sub-Jovian based on WASP-107 \citep{Macleod2022}. Using the same input parameters as were used in the simulations, we calculate the trajectory of the tail using our model, shown by the thick black line. The model shows good agreement with the simulations in all cases.}
\label{fig:Comparison_to_hydro}
\end{figure*}

Figure~\ref{fig:Comparison_to_hydro} shows the trajectory of the tail calculated using our model superimposed on the outflow produced from 3D hydrodynamics simulations. Our model is successful in reproducing the trajectory of the simulated tails in all the simulations, and this gives us confidence that it can be used to test atmospheric escape models quantitatively.  

\section{Computing Synthetic Transit Profiles}
\label{sec:Lightcurve}

The primary motivation of this model is to produce synthetic transmission spectra that can be compared to observations. We use a ray-tracing scheme to calculate the attenuation of Lyman-$\alpha$ due to the transit of the outflow over the stellar disc. We produce observations at different orbital phases by rotating the outflow around the system's barycenter, which we approximate as the star's centre. To compute the attenuation of Lyman-$\alpha$ at a particular orbital phase, we create a 3D cylindrical grid from the product of a 2D grid on the stellar disc and a 1D grid along the line connecting the centre of the stellar disc to the observer. We then define a new cartesian coordinate system with the origin at the centre of the stellar disc. The $x_t$ and $y_t$ axes are parallel to the stellar disc, and the $z_t$ axis is given by the line connecting the centre of the stellar disc to the observer. The 2D grid consists of 705 cells. For ray tracing through the tail, the 1D grid consists of a minimum of 10 cells across the width of the tail. For ray tracing through the Hill sphere, the 1D grid consists of 15 cells. We did resolution tests in order to check that this number of cells was sufficient.  

The optical depth through the outflow for a corresponding point on the stellar disc is given by

\begin{align}
\label{eq:los_optical_depth}
\tau_{\nu}(x_t,y_t) = \int_{0}^{\infty}n_{\text{HI}}(x_t, y_t, z_t)\sigma_{\nu}(u_{z_t}, T)dz_t
\end{align}

\noindent where $n_{\text{HI}}$ is the number density of hydrogen atoms, $\sigma_{\nu}$ is the absorption cross-section of hydrogen to photons of frequency $\nu$, which is a function of the line of sight velocity of the gas and the temperature. This accounts for Lyman-$\alpha$ optical depth due to both neutral hydrogen of planetary origin and ENAs of stellar origin.  

The absorption cross-section is given by

\begin{align}
    \sigma_{\nu} = \frac{\pi e^2}{m_{e}c}f\Phi(u_{z_t}, T)
\end{align}

\noindent where $e$ and $m_e$ is the charge and mass of the electron, $c$ is the speed of light, $f = 4.1641 \times 10^{-1}$ is the oscillator strength of the transition and $\Phi$ is the Voigt line profile \citep{NIST_ASD}. The Gaussian part of the Voigt profile has a standard deviation of $\sigma_N$, the Lorentzian part has a half-width at half-maximum of $\gamma$, and the line centre has been Doppler shifted to $\nu$. These are given below:

\begin{align}
    \sigma_N = \sqrt{\frac{k_bT}{m_{H} c^2}}\nu_0 ,\quad \gamma = \frac{A}{4\pi}, \quad \nu = \nu_0\left(1 - \frac{u_{z_t}}{c}\right)
\end{align}
where $k_b$ is the Boltzmann constant. Here $\nu_0$ is the rest frequency of Lyman-$\alpha$ and $A = 6.2649 \times 10^{8} \text{ s}^{-1}$ is the Einstein $A$ coefficient of the transition \citep{NIST_ASD}. 

The obscuration fraction averaged over the stellar disc is: 

\begin{align}
\label{eq:integrate_stellar_disc}
\mathcal{O}_{\nu} = \frac{\iint_{D}e^{-\tau_{\nu}(x_t,y_t)}dx_tdy_t}{\pi R_{*}^2}
\end{align}
Here, we do not account for spatial variations in the Lyman-$\alpha$ profile across the stellar surface, as seen on the Sun \citep[][]{Gordino2022}. To numerically solve for the obscuration fraction, the integral in Equation~\ref{eq:los_optical_depth} is calculated by performing a middle Riemann sum using the 1D grid in the $z_t$-direction. The integral in Equation~\ref{eq:integrate_stellar_disc} is calculated by performing a double Riemann sum using the 2D grid over the stellar disc \citep[e.g.,][]{Schreyer2024}.  

\section{Comparison to Real Systems}
\label{sec:fitting_procedure}

In order to assess the biases in our model, we check the ability to recover known parameters from mock observations. For a parameter set $\boldsymbol{\Theta}$, we calculate the obscuration due to the planetary outflow as a function of time and wavelength. We bin the resulting spectrum into 30-minute time bins and three velocity bins between [-150, -50] km s$^{-1}$, and add error bars $\sim 10 \%$ to the resulting points. We make twenty realisations of the synthetic data by re-sampling the points randomly within its error. We perform a retrieval on each of these realizations. For each, we are able to recover the true parameters within 1$\sigma$, and the combination of the twenty realizations was not biased.   

\subsection{Fitting the transit of GJ 436 b}

We use the outflow model to perform a retrieval of the warm Neptune GJ 436 b in a Bayesian framework. We choose this planet because the Lyman-$\alpha$ transit of this planet has been extensively observed with the Hubble Space Telescope \citep{Kulow2014, Eherenreich2015, Lavie2017, dosSantos2019}. The fixed and free parameters of the model are listed in Table~\ref{tab:Fixed and Free Parameters}. Here, we list the values used for the fixed parameters and the prior bounds for the free parameters. In order to reduce the number of free parameters, we set the temperature of the stellar wind to 0.5 MK. We checked that this choice does not affect our conclusions by running models with $T_{\text{sw}} = 1$ MK, in which we found similar results. In the fitting procedure, we convert the angular momentum deficit, $h$, into the "launch angle" of the outflow, given by $\theta$ in \ref{eq:ang mom def}. We do this so that we can put a uniform prior on the launch angle. For all other free parameters, apart from the inclination, we use log uniform priors, as the scale of these parameters is not constrained. The prior bounds enclose realistic parameter bounds as determined from theory or other observations. We use the energy-limited mass loss rate, 

\begin{align}
\label{eq:el_mlr}
\dot{M}_{\text{EL}} = \eta\frac{\pi F_{\text{XUV}}R_P^3}{GM_p}
\end{align}
with the efficiency $\eta = 1$ to set an upper bound on the planetary mass loss rate. Here, the EUV flux is approximated from the optically thin photoionization rate by assuming all ionizing photons deliver $20 \text{ eV}$ (see \citet{Murray-Clay2009}, Appendix~\ref{appendix2}). We note that the appropriate radius over which the planet absorbs EUV flux may be larger than the optical radius, which can enhance this mass loss rate limit. However,  this is often balanced out by the real efficiency for escaping planets being significantly less than one. Therefore, in our case, it is reasonable to simply use the optical radius.   

\begin{table}
  \label{tab:Fixed and Free Parameters}
  \caption{The values of the fixed parameters and the priors on the free parameters used in fitting the Lyman-$\alpha$ transit of GJ 436 b}
  \begin{tabular}{p{6.5cm}p{1.5cm}}
   Parameter & Value/Prior\\
   \hline
   Fixed\\
   \hline
   Planet Mass, \ $M_p$ ($M_J$) & 0.07\\
   Planet Radius, \ $R_p$ ($R_J$) & 0.35\\
   Semimajoraxis, \ $a_p$ (AU) & 0.029\\
   Stellar Mass, $M_{*}$ ($M_{\odot}$) & 0.45\\
   Stellar Radius, $R_{*}$ ($R_{\odot})$ & 0.425\\
   Stellar Wind Temperature, $T_{\text{sw}}$ (MK) & 0.5 \\
   \hline
   Free\\
   \hline
   Log Sound speed (cm s$^{-1}$) & U(5.2, 6.5)\\ 
   Log Planet mass loss rate (g s$^{-1}$) & U(8, $\dot{M}_{\text{EL}}$) \\
   Log Stellar wind velocity (at planet) (cm s$^{-1}$) & U(6.5, 8)\\
   Log Star mass loss rate (g s$^{-1}$) & U(10.3, 13)\\
   Log Optically thin photoionization rate (at planet) (s$^{-1}$) & U(-5.6, -2.6)\\
   Launch Angle (radians) & U($\frac{\pi}{2}$, $\pi$)\\
   Inclination (radians) & N(1.51, 0.02)\\
   Log Bulk velocity of ENAs (cm s$^{-1}$)  & U(6.4, $u_{*}$) \\
   Log Width of turbulent mixing layer & U(-2, -0.4)\\
   \end{tabular}
\end{table}

We sample the posterior using the Markov chain Monte Carlo (MCMC) ensemble sampler \textit{emcee} \citep{ForemanMackey2013}. This is a Python implementation of Goodman \& Weare's affine invariant MCMC ensemble sampler \citep{Goodman2010}. We calculate the log-likelihood by comparing the observed flux $\mathcal{F}_{o}$ to the modelled flux $\mathcal{F}$ as functions of both time and wavelength within a collection of observed time series spectra as:  

\begin{align}
\text{ln}\mathcal{L}(\boldsymbol{\Theta}) = -\frac{1}{2}\sum_{n}\left[\frac{(\mathcal{F}_{o} - \mathcal{F})^2}{\sigma^2_{n}} + \text{ln}(2\pi \sigma^2_{n})\right] 
\label{eq:likelihood}
\end{align}

The time-series spectra consist of all Lyman-$\alpha$ observations of GJ~436 to date taken with the Hubble Space Telescope's (HST's) Space Telescope Imaging Spectrograph (STIS) configured with the G140M grating. This consists of 26 individual observations obtained between 2010 and 2016 (program PIs Ehrenreich, Green, and France). We uniformly re-extract the spectra using the calSTIS pipeline, manually identifying the position of the spectral trace on the detector.  We use an extraction ribbon of 11 pixels and background ribbons of 20 pixels at $\pm$20 pixels from the trace location to minimize effects from a spatially-variable background systematic known as FUV glow. The observations are divided into subexposures of roughly 500 s each, with minor adjustments to ensure the full use of each exposure. 

For comparison to the data, we forward-model the observed Lyman-$\alpha$ spectrum during each subexposure. We do this by applying the absorption predicted by the model ray tracing to a de-broadened version of the stellar Lyman-$\alpha$ profile as observed at earth reconstructed using the technique described in \cite{wilson2022}.  The profile is then broadened by the instrument line spread function and binned to match the wavelength grid of the data. The observed and modelled spectra are then normalized to an identical flux in the range of 90 and 250 km s$^{-1}$. This accounts for known variability in instrument throughput and stellar variability in which the full spectrum is scaled by a constant. We note that this method cannot take into account variability in which the ratio of flux in the red wing to the blue wing changes. After normalization, we use the predicted spectrum and flux scale of the instrument to estimate the expected number of counts in each wavelength bin, then use this to compute the expected variance due to Poisson statistics. This mitigates problems due to zero-flux, zero-error points in the calSTIS-reduced data and the disproportional influence of low-flux, low-error points on fits and is more appropriate for our Bayesian framework. We use the resulting normalized spectra and estimated errors to compute the data likelihood with Equation~\ref{eq:likelihood}. We comment that there are minor differences in the transit depth in our reduced data and that of previous authors \citep{Lavie2017} as a result of our different normalization strategy, and possibly different transit baseline.

We ran the MCMC for a minimum of 15000 steps with 100 walkers. In order to monitor convergence, we ran two independent chains for each model so that we could calculate the split chains potential scale reduction factor, split-$\hat{R}$ \citep[][]{Gelman2014}. The maximum value across all parameters was 1.02. We also calculated the effective sample size for each of the parameters, which was a minimum of 15. Based on these results, we conclude that the posterior we have calculated from the MCMC is representative of the true posterior. 

\section{Results and Discussion}
\label{sec:results}

To explore how including different parts of the outflow affects inferences, we run four models of increasing complexity. In the first, we only include the tail outflow; in the second, we include both the tail and the hill sphere outflow; in the third, we include the tail outflow and ENAs; and in the fourth, we include all model components (i.e. the tail, hill sphere outflows, and ENAs). In the models where we do not include the outflow inside the Hill sphere, we only compare to data after the egress of the Hill sphere ($> 1.3 \text{ hours}$). In the cases where we include the outflow inside the Hill sphere, we only compare to data after the egress of the planet ($> 0.7 \text{ hours}$). The observed transit shows absorption before the ingress of the Hill sphere, and therefore we avoid comparing to data affected by the outflow in front of the Hill sphere of the planet, where we underestimate the transit depth. The retrieved parameters from all four models are consistent, indicating the tail part of the outflow drives our inferences. This is expected for GJ 436 b as the tail dominates the transit duration. We find that ENAs contribute very little to the Lyman-$\alpha$ optical depth. For the retrieved parameters, the mixing region is $\lesssim$ 3 percent of the depth of the tail $\sim R_p$, which leads to very small optical depths (< 0.01) at typical ENA densities on the order of the stellar wind density. We verify this by calculating the spectrum for our best-fit model without the contribution of ENAs and find it virtually indistinguishable from the spectrum including ENAs.  

We do find a slight inconsistency in retrieved inclinations between the models with and without the inclusion of the Hill sphere. In the cases with the Hill sphere, the inclination has a median of 1.55 radians, whilst in the cases without the Hill sphere, it has a median of 1.52 radians, more in line with observational constraints. We think this discrepancy comes from the fact that cutting the gas off at the Hill sphere underpredicts the extent of the gas in this region, and therefore, the inclination has to change to compensate. However, the fact the rest of the retrieved parameters are the same gives us confidence that most of the information is coming from the tail. As a result, we focus on the full model from now on. 

\begin{figure*}
\centering
\includegraphics[width=0.47\textwidth]{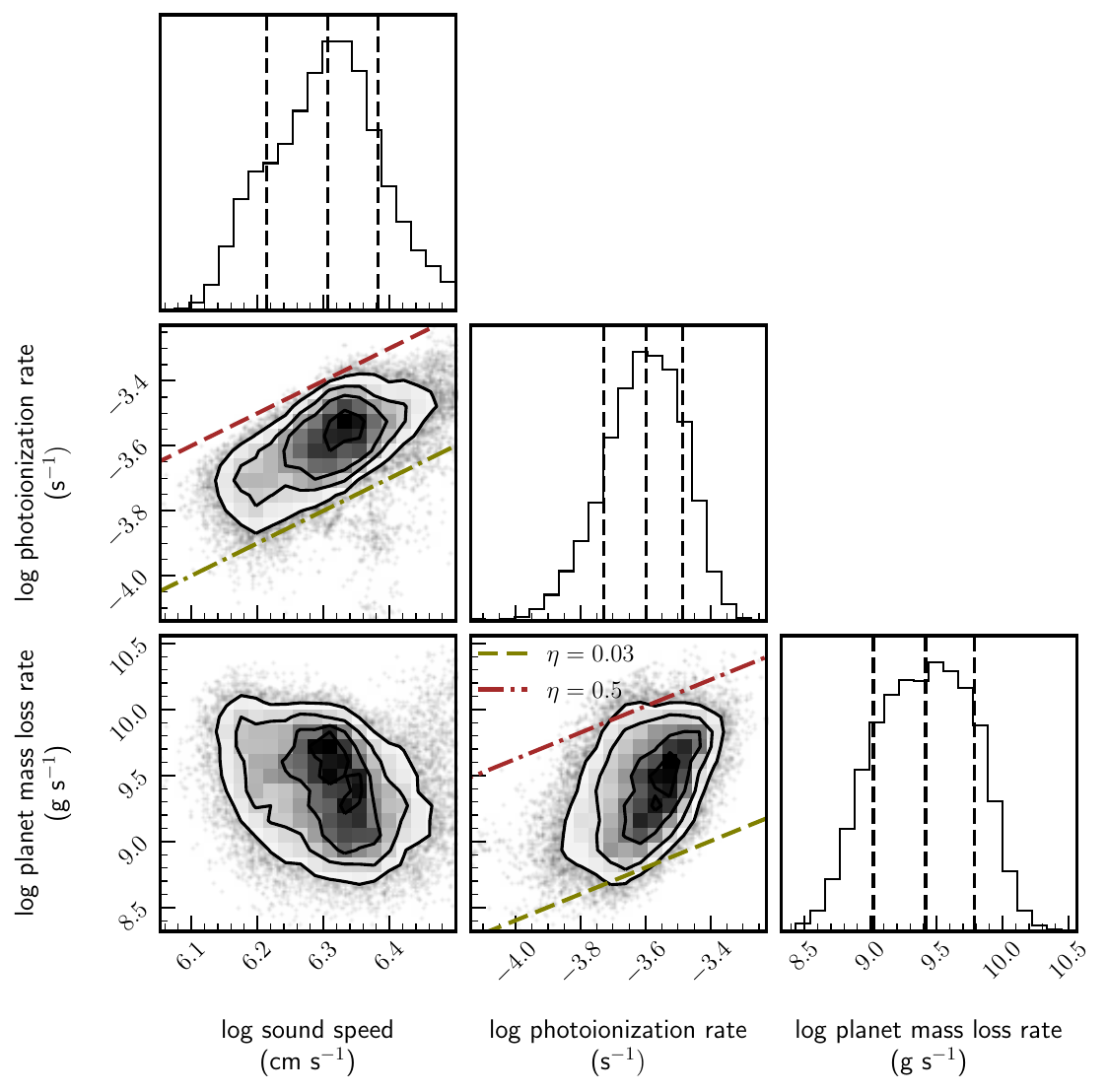}
\includegraphics[width=0.47\textwidth]{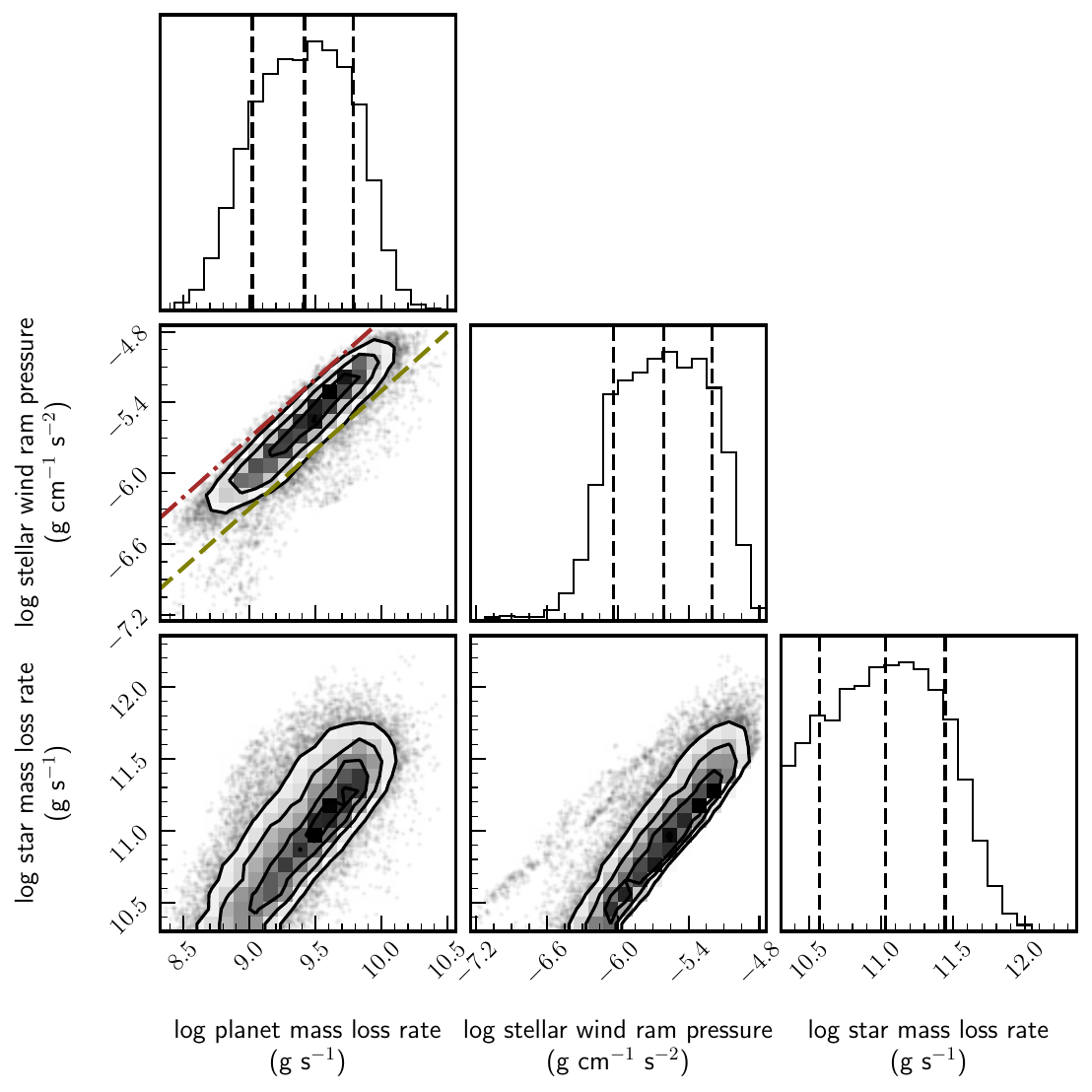}
\caption{Left: Posterior distribution of the sound speed, planet mass loss rate, and photoionization rate for GJ 436 b. On the joint sound speed and photoionization rate posterior, we have plotted lines of constant ratio corresponding to $\frac{c_s}{\Gamma_p} = 10^{9.7-10.1}$ cm. On the joint mass loss rate and photoionization rate posterior, we have plotted lines of constant efficiency from the energy-limited mass loss formula (Equation~\ref{eq:el_mlr}). Right: Posterior distribution of the planet mass loss rate, stellar wind ram pressure, stellar mass loss rate. On the joint planet mass loss rate and stellar wind ram pressure posterior, we plot lines of constant ratio corresponding to $\frac{\dot{M}_{*}u_{*}}{\dot{M}_{p}} = 10^{9.1-9.7} \text{ cm s}^{-1}$. Although planetary mass loss rate and stellar wind ram pressure cannot be tightly constrained individually, the tight correlation between these parameters means that this ratio is tightly constrained, illustrated by the small spacing of constant-ratio lines bounding the posterior.}
\label{fig:posterior}
\end{figure*}

\begin{figure*}
\centering
\includegraphics[width=\textwidth]{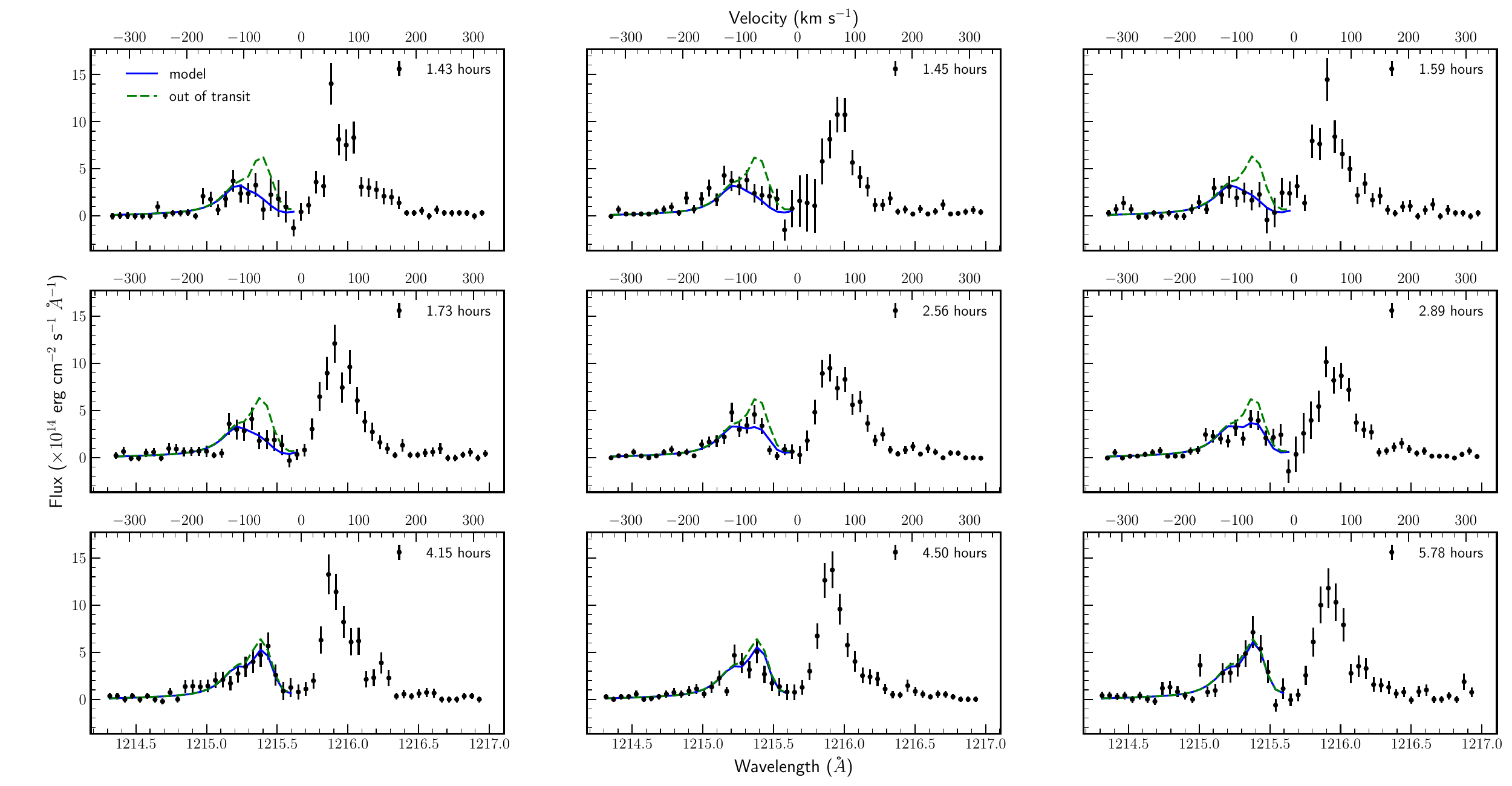}
\caption{Lyman-$\alpha$ transit spectrum of GJ 436 b at different times (relative to optical mid-transit) measured using Hubble Space Telescope (black points) and the model (solid blue) using the MAP estimates from the MCMC sampler. The dashed green line is the out-of-transit model that is used to calculate the transmission spectrum.}
\label{fig:transmission_spectra}
\end{figure*}

\begin{figure}
\centering
\includegraphics[width=0.48\textwidth]{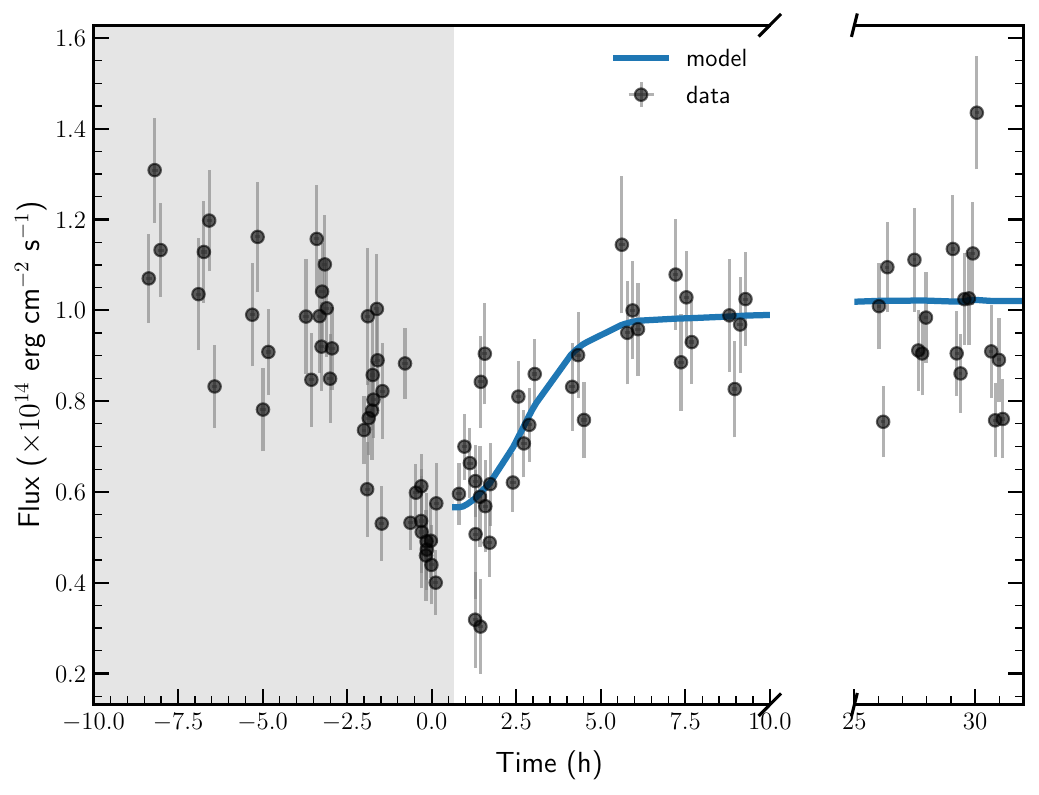}
\caption{Lyman-$\alpha$ transit of GJ 436 b integrated over the blue wing interval of [-150, -50] km s$^{-1}$. The shaded region is before the egress of the planet, where gas leading the Hill sphere of the planet is transiting the star, which we do not fit with our model.}
\label{fig:lightcurve}
\end{figure}

In Figure~\ref{fig:posterior}, we present the marginalised posterior distributions for the sound speed, planetary mass loss rate, optically thin photoionization rate, mass loss rate of the star and ram pressure of the stellar wind. We provide the full posterior distribution in Appendix~\ref{appendix1}. The 2D joint posteriors show correlations between the parameters, which help elucidate the key physics governing observations. These correlations are easier to extract in this framework than from only running 3D simulations because our model is significantly less computationally expensive than 3D simulations so is able to more widely sample the possible parameter space. Later in this section, we examine how the retrieved sound speed of the escaping gas, planetary mass loss rate and photoionization rate in the tail, which control the density of neutral hydrogen, are correlated. We also examine how the retrieved planetary mass loss rate and stellar wind ram pressure, which control the radial velocity of the outflow, are correlated. In Figure~\ref{fig:transmission_spectra}, we plot the model transmission spectra at different times in the transit for the maximum a posteriori parameter estimates (i.e. posterior mode). It is clear that the model is a good match for the spectrum, providing evidence that the model accurately represents the planetary outflow. This is further illustrated by Figure~\ref{fig:lightcurve}, where we show the transit integrated between -50 and -150 km s$^{-1}$. 

\subsection{Mass Loss Rate}

We find a total hydrogen mass loss rate of $2.6^{+3.5}_{-1.6} \times 10^9 \text{ g s}^{-1}$. This is similar to the mass loss rate of $2\times10^{9}$~g~s$^{-1}$ found in the 1D simulations by \citet{Schulik2023}, and $5.5 \times 10^9 \text{ g s}^{-1}$ found in 3D hydrodynamics simulations by \citet{Villarreal2021}, which were able to reproduce the depth and duration of the transit. Considering the joint mass loss rate and photoionization, we can also put constraints on the efficiency of the outflow, $\eta$, given in the energy-limited formula (Equation~\ref{eq:el_mlr}). On the posterior, we plot lines of constant mass loss efficiency and are able to constrain this to range from 0.03-0.5. This is consistent with simulations of atmospheric escape from GJ 436 b that accounted for both EUV and bolometric heating and found an efficiency of $\sim 0.05$ \citep{Schulik2023}. More generally, it is consistent with the theoretical estimates for a photoevaporative flow \citep[e.g.][]{Owen2012, Owen2016}. We highlight that as a real stellar spectrum has significant flux upwards of 20 eV, these efficiencies are likely underestimated by a factor of a few (see Appendix~\ref{appendix2}). 

The slope of the correlation between mass loss rate and flux is much weaker than the energy-limited slope. Even though higher EUV fluxes lead to a higher mass loss rate, the outflow is more ionized to the extent that the density of neutral hydrogen actually decreases. Therefore, in order to match the observed optical depth to Lyman-$\alpha$ photons from neutral hydrogen, the model has to compensate for a higher EUV flux with more mass loss relative to the energy-limited relationship.    

\begin{figure}
\centering
\includegraphics[width=0.48\textwidth]{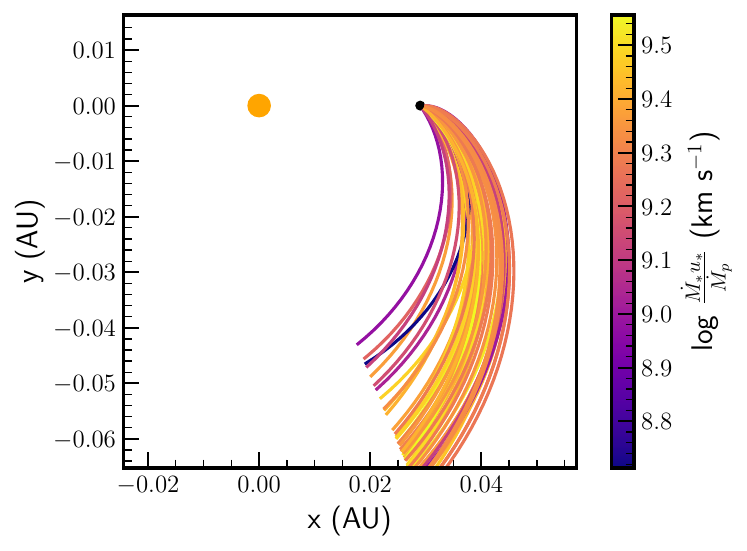}
\caption{A sample of tail trajectories drawn from the posterior distribution from the retrieval of GJ 436 b coloured by the ratio of stellar wind ram pressure to planetary mass loss rate. Whilst we note that the trajectory of the tail weakly depends factors such as the launch angle and outflow sound speed, the ratio of the stellar wind pressure to planetary mass loss rate controls the acceleration of the tail, and therefore is primarily responsible for setting the trajectory of the tail.}
\label{fig:tails}
\end{figure}

Interestingly, we put a tight constraint on the ratio of the ram pressure of the stellar wind at the planet and the mass loss rate of the planet of $\frac{\dot{M}_{*}u_{*}}{\dot{M}_p} = 10^{9.4 \pm 0.3} \text { cm s}^{-1}$. The reason for this is that this ratio determines the radial acceleration of the tail and, hence the radial velocity and the position of the tail as a function of time. In Figure~\ref{fig:tails}, we plot a sample of tail trajectories drawn from the posterior distribution from the retrieval of GJ 436b, coloured by the ratio of the stellar wind ram pressure to the mass loss rate of the planet. This demonstrates the dominant role that this ratio plays in setting the trajectory of the outflow and the sensitivity of the observed transit to stellar wind conditions.

To see the effect that the acceleration of the tail has on the observation spectrum, we consider the ratio of observed flux in the (-150,-100) km s$^{-1}$ band and (-100,-50) km s$^{-1}$ band, shown in Figure~\ref{fig:flux_ratio}. As the tail is accelerated, more flux is absorbed at higher velocities, so the ratio decreases. 
We note that this ratio also changes as the column density of neutral hydrogen changes, and therefore, to extract the effect of the acceleration of the tail, we have shown how this ratio evolves if the gas in the tail was fully neutral.       

We highlight that acceleration of the tail (Equation~\ref{eq:tail_acceleration}) is not exactly proportional to the ram pressure of the stellar wind but depends on the difference between the radial velocity of the tail and the stellar wind. The maximum velocity that the tail can be accelerated to is the stellar wind velocity. If the tail reaches this terminal velocity whilst still attenuating Lyman-$\alpha$ photons, then very strong constraints can be placed on the stellar wind velocity. In the case of GJ 436 b, the tail is still accelerating, and therefore, we are only able to put a lower limit of ~300 km s$^{-1}$ on the stellar wind velocity at the planet. 

We emphasize that we do not dismiss the case where the stellar wind velocity is very high (> 800 km s$^{-1}$), and the mass loss rate is low as GJ 436 b has a polar orbit \citep[][]{Bourrier2022}. For the Sun, the polar wind is significantly faster and less dense than the equatorial wind, and therefore, it is possible that this is the case for GJ 436 as well.

\subsection{Sound Speed of the Outflow}
\label{sec:ss}

The $2 \sigma$ lower bound for the retrieved sound speed is $13 \text{ km s}^{-1}$, corresponding to a temperature of $\sim$ 10, 000 K for a fully ionized purely hydrogen gas, which is consistent with a photoevaporative outflow. The bulk of the posterior encompasses higher sound speeds, with a MAP estimate of $20.5 \text{ km s}^{-1}$, which corresponds to outflow temperature $\sim$ 40,000 K. This is higher than expected for photoevaporative outflows which are usually thermostat at temperatures $\sim$ 10,000 K, due to strong radiative cooling of hydrogen above those temperatures. 
 
It is possible that another physical mechanism is accelerating the outflow to high velocities by the time it reaches the Hill sphere, raising the apparent sound speed. This interpretation is compatible with the results of \citet{Bourrier2016}, which used a collisionless outflow model of the exosphere, and required outflow velocities of $\sim 50-60 \text{ km s}^{-1}$ in order to fit the data. A possible mechanism to produce these high velocities is via the dissipation of MHD waves in the upper atmosphere \citep[e.g][]{Tanaka2014}; however, consistent multi-dimensional models are needed to prove that this is a viable mechanism to drive high-velocity outflows.

Another possible reason for the high retrieved sound speed is that our method underestimates the production of ENAs, which provide a source of high velocity hydrogen atoms. This is particularly important during the Hill sphere transit, where the velocity of the planetary gas blocking the star is (relatively) low. To compensate, the optical depth of the planetary outflow in the Lyman-$\alpha$ blue wing needs to be increased. This can be achieved by increasing the sound speed of the outflow. We note that single-fluid model for charge exchange production predicts a significantly lower contribution to ENAs compared to particle based \citep[e.g.][]{Bourrier2016} or multi-fluid hydrodynamic simulations \citep[e.g.][]{Khodachenko2019}, which are able to reproduce large mid-transit depths with Lyman-$\alpha$ absorption from ENAs.  

To assess whether our single-fluid model for the production of ENAs is consistent, we check that conditions for a single-fluid framework hold. In our best fit model, the density of protons and hydrogen atoms at the ionopause (located $\sim 7.5 R_p-1.8 \times 10^{10} \text{ cm}$), is $\sim 10^5 \text{ cm}^{-3}$. At $10^6 \text{ K}$, the cross section for proton-hydrogen collisions is $\sim 6 \times 10^{-15} \text{ cm}^{2}$ \citep{Schultz2008} and the cross section for proton-proton collisions is $\sim 10^{-16} \text{ cm}^2$ \citep{Schunk1980}. Therefore the mean free path for stellar wind protons in the planetary outflow is $\sim 10^9 \text{ cm}$. This is 18 times smaller than the ionopause radius, therefore it is valid to use the single-fluid framework. We comment that in different parts of the parameter space (for example, at lower planetary wind densities), this condition may not hold, therefore particle-based or multi-fluid methods are required to properly model the production of ENAs. 

One potential way to break the degeneracy between the contribution of the planetary outflow and ENAs is to consider the distinction between the velocity-resolved transit from each of these factors. If the planetary wind is responsible for the majority of Lyman-$\alpha$ optical depth, then the velocity of peak absorption should increase over the duration of the transit as the tail is accelerated away from the star. ENAs, which are of stellar origin, will have a different velocity distribution and therefore the observed velocity evolution of the transit may differ from the that produced by hydrogen atoms of planetary origin. To determine if this is the case, more work investigating the velocity structure of generated ENAs and how they imprint upon velocity-resolved transits as a function of planetary and stellar parameters is needed.    

\begin{figure}
\centering
\includegraphics[width=0.48\textwidth]{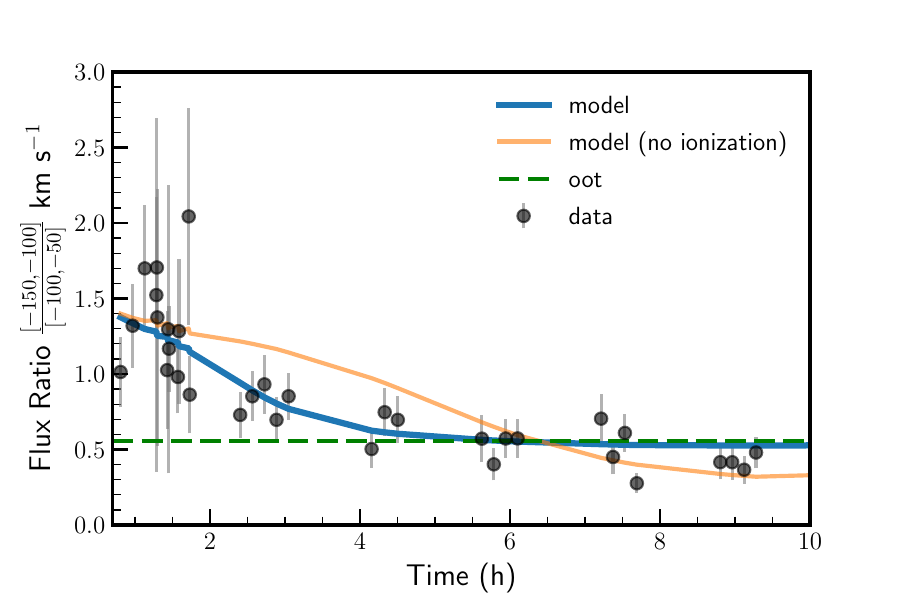}
\caption{The observed flux ratio between (-150, -100) and (-100, -50) km s$^{-1}$ as a function of time after the transit (black dots). The blue line is the best-fit model, the orange line is the best-fit model, setting the photoionization rate to zero, and the dashed green line is the out-of-transit ratio. There is a smooth decrease in this ratio from $\sim$ 1 hour to 5 hours as the tail is accelerated to higher velocities and becomes more ionized, absorbing more photons in the (-150, -100) km s$^{-1}$ band compared to the (-100, -50) km s$^{-1}$ band. The model with the ionization rate set to zero indicates the length of the transit required in order to unambiguously detect the acceleration of the tail (as ionization also changes the flux ratio).}
\label{fig:flux_ratio}
\end{figure}

We note that in the case of GJ 436 b, we cannot definitively tell that the absorbing material is being accelerated as a function of time (model independently) because the gas becomes ionized too quickly for the velocity of peak absorption to change dramatically (see Figure~\ref{fig:flux_ratio}). However, for planets with lower ionization rates and, therefore, longer tails (e.g. K2-18b, \citealt{dosSantos2020}) or when the gas in the tail is in ionization-recombination equilibrium, it may be possible to observe the acceleration of the gas in the lightcurve.

\subsubsection{Limitations and Future Improvements}

Since we de not model the gas in front of the Hill sphere of the planet, we are not able to account for the information about the outflow in the pre-transit data. The challenge with modelling the outflow leading the planet is that it is unstable, resulting in the variable occultation of the star \citep[e.g.,][]{McCann2019} as seen in the case of AU Mic b \citep{Rockcliffe2023}. The reason for this instability is that the Coriolis force and the stellar wind pressure both act in the same direction; therefore, the outflow cannot be in generalised geostrophic balance and reach a stable density structure. This problem is only pertinent when the planetary wind and the stellar wind have similar strengths. However, Lyman-$\alpha$ observations, which show an extended outflow leading the planet \citep[e.g.,][]{Rockcliffe2023} make understanding this case important. We leave modelling of the outflow leading the planet to future work. In the case of GJ 436 b, where there is minimal pre-transit absorption, the stellar wind is most likely strong enough to stop a leading outflow. In this case, the time of ingress provides an approximate location for the boundary between the stellar and planetary wind, which can be used to infer their relative pressures at this location. 

Our model does not include the effect of magnetic fields, as we use purely hydrodynamic simulations as the basis for our work. Both the planetary magnetic field and the stellar wind magnetic field can affect the geometry of the outflow \citep[e.g.,][]{Carolan2021b}. However, it is not yet clear whether close-in exoplanets host strong magnetic fields \citep[e.g.,][]{Schreyer2024} or whether the stellar wind magnetic field plays an important role in the interaction with the planetary outflow as the ram pressure of the stellar wind dominates over the magnetic pressure outside the Alfvén surface.    

Similarly, we do not include the effects of radiation pressure. Like the stellar wind, this can impart momentum to the planetary outflow, accelerating it radially. For GJ 436 b, radiation pressure is unlikely to be important \citep[e.g.,][]{Khodachenko2019}. However, this may not be the case for all systems. In the future, we will extend the model to include a term accounting for this force. Another possible improvement to this model would be to include a more realistic stellar wind accounting for the orbital velocity of the planet and variations along the orbit.    

We also imposed a circular orbit on the planet. Since highly irradiated planets are close to their host star, they are generally assumed to have undergone orbital circularization. However, deviations from a circular orbit can cause a line of sight velocity shift of $v_ke\text{cos}\omega$, where $e$ is the eccentricity of the planet and $\omega$ is the argument of periastron. In the case of GJ 436 b, the eccentricity and argument of periastron are 0.156 and 325.8 degrees, respectively \citep[][]{Bourrier2022}, which equates to a line of sight velocity shift of $\sim -15 \text{ km s}^{-1}$ at mid-transit. This is small compared to the velocities probed by Lyman-$\alpha$ transits and, therefore, cannot be fully responsible for the high radial velocities observed. However, it may partially be responsible for the higher-than-expected sound speed retrieved. In the future, it may be worthwhile folding constraints on the planetary orbital dynamics into the model. 

\section{Conclusions}

We have created a model of the interaction of a planetary outflow with the circumstellar environment, which can be used to perform synthetic Lyman-$\alpha$ transits. We demonstrate that this model is able to reproduce the geometry of outflows produced by 3D hydrodynamics simulations. Furthermore, we illustrate the usage of this model to perform the first retrieval of atmospheric escape properties using velocity-resolved Lyman-$\alpha$ observations. 

For GJ 436 b, we are able to constrain the sound speed of the escaping gas to be $ \gtrsim 10 \text{ km s}^{-1}$, indicating that we can rule out core-powered mass loss as the mechanism of escape for this planet. This is consistent with the predictions from the combined photoevaporation and core-powered mass-loss model of \citet{Owen2024}. Although the retrieved planetary mass loss rate has wide bounds, we find that the ratio of planetary mass loss rate to stellar wind ram pressure is tightly constrained. Independent constraints on the space environment around this planet \citep[e.g.,][]{Bellotti2023} would mean that we are able to put a stronger constraint on the mass loss rate. Equally, as we expect this ratio to be tightly constrained if the tail is observable, we can also use Lyman-$\alpha$ transits as a way to measure the stellar winds of stars.  

The usefulness of the simple model is that we are able to efficiently explore the large parameter space to find solutions that match the data well. In doing so, we are able to put constraints on parameters and find correlations between parameters, which can both inform whether current models can fit the observations and guide future 3D simulations. We do caution that making a simplified model requires sacrificing some accuracy as compared to 3D simulations, and therefore, further simulations are needed to confirm the conclusions drawn from this study. At the same time, although much progress has been made with large-scale simulations, we also note the physics included, and hence, the validity to describe Lyman-$\alpha$ transits is still debated. Therefore, simple models that can provide tests of the physics of Lyman-$\alpha$ transits on both individual systems and on a population-wide basis are essential to progress in this field.

\section*{Acknowledgements}

We thank the anonymous reviewer for their comments, which improved the manuscript. JEO is also supported by a Royal Society University Research Fellowship.  ES \& JEO. has also received funding from the European Research Council (ERC) under the European Union’s Horizon 2020 research and innovation programme (Grant agreement No. 853022, PEVAP). The authors acknowledge support from NASA XRP grant 80NSSC23K0282.   This research is based on observations made with the NASA/ESA Hubble Space Telescope obtained from the Space Telescope Science Institute, which is operated by the Association of Universities for Research in Astronomy, Inc., under NASA contract NAS 5–26555. These observations are associated with program(s) 11817, 12034, 12965, 13650, and 14222. This work benefited from the 2022 Exoplanet Summer Program in the Other Worlds Laboratory (OWL) at the University of California, Santa Cruz, a program funded by the Heising-Simons Foundation. For the purpose of open access, the authors have applied a Creative Commons Attribution (CC-BY) license to any Author Accepted Manuscript version arising.

\section*{Data Availability}

The data used in this work are publicly available through the Barbara A. Mikulski Archive for Space Telescopes and have been aggregated under DOI 10.17909/pjvt-h370\footnote{\url{http://dx.doi.org/10.17909/pjvt-h370}}. The code used to generate these results is available on \href{https://github.com/eschreyer/LyA_code}{github}. 



\bibliographystyle{mnras}
\bibliography{Bibliography} 

\begin{thebibliography}{}
\makeatletter
\relax
\def\mn@urlcharsother{\let\do\@makeother \do\$\do\&\do\#\do\^\do\_\do\%\do\~}
\def\mn@doi{\begingroup\mn@urlcharsother \@ifnextchar [ {\mn@doi@} {\mn@doi@[]}}
\def\mn@doi@[#1]#2{\def\@tempa{#1}\ifx\@tempa\@empty \href {http://dx.doi.org/#2} {doi:#2}\else \href {http://dx.doi.org/#2} {#1}\fi \endgroup}
\def\mn@eprint#1#2{\mn@eprint@#1:#2::\@nil}
\def\mn@eprint@arXiv#1{\href {http://arxiv.org/abs/#1} {{\tt arXiv:#1}}}
\def\mn@eprint@dblp#1{\href {http://dblp.uni-trier.de/rec/bibtex/#1.xml} {dblp:#1}}
\def\mn@eprint@#1:#2:#3:#4\@nil{\def\@tempa {#1}\def\@tempb {#2}\def\@tempc {#3}\ifx \@tempc \@empty \let \@tempc \@tempb \let \@tempb \@tempa \fi \ifx \@tempb \@empty \def\@tempb {arXiv}\fi \@ifundefined {mn@eprint@\@tempb}{\@tempb:\@tempc}{\expandafter \expandafter \csname mn@eprint@\@tempb\endcsname \expandafter{\@tempc}}}

\bibitem[\protect\citeauthoryear{{Allart} et~al.,}{{Allart} et~al.}{2018}]{Allart2018}
{Allart} R.,  et~al., 2018, \mn@doi [Science] {10.1126/science.aat5879}, \href {https://ui.adsabs.harvard.edu/abs/2018Sci...362.1384A} {362, 1384}

\bibitem[\protect\citeauthoryear{{Bellotti} et~al.,}{{Bellotti} et~al.}{2023}]{Bellotti2023}
{Bellotti} S.,  et~al., 2023, \mn@doi [arXiv e-prints] {10.48550/arXiv.2306.15391}, \href {https://ui.adsabs.harvard.edu/abs/2023arXiv230615391B} {p. arXiv:2306.15391}

\bibitem[\protect\citeauthoryear{{Ben-Jaffel} et~al.,}{{Ben-Jaffel} et~al.}{2022}]{Ben-Jaffel2022}
{Ben-Jaffel} L.,  et~al., 2022, \mn@doi [Nature Astronomy] {10.1038/s41550-021-01505-x}, \href {https://ui.adsabs.harvard.edu/abs/2022NatAs...6..141B} {6, 141}

\bibitem[\protect\citeauthoryear{{Bisikalo}, {Kaygorodov}, {Ionov}, {Shematovich}, {Lammer}  \& {Fossati}}{{Bisikalo} et~al.}{2013}]{Bisalko2013}
{Bisikalo} D.,  {Kaygorodov} P.,  {Ionov} D.,  {Shematovich} V.,  {Lammer} H.,   {Fossati} L.,  2013, \mn@doi [\apj] {10.1088/0004-637X/764/1/19}, \href {https://ui.adsabs.harvard.edu/abs/2013ApJ...764...19B} {764, 19}

\bibitem[\protect\citeauthoryear{{Bourrier} \& {Lecavelier des Etangs}}{{Bourrier} \& {Lecavelier des Etangs}}{2013}]{Bourrier2013b}
{Bourrier} V.,  {Lecavelier des Etangs} A.,  2013, \mn@doi [A\&A] {10.1051/0004-6361/201321551}, 557, A124

\bibitem[\protect\citeauthoryear{{Bourrier} et~al.,}{{Bourrier} et~al.}{2013}]{Bourrier2013}
{Bourrier} V.,  et~al., 2013, \mn@doi [\aap] {10.1051/0004-6361/201220533}, \href {https://ui.adsabs.harvard.edu/abs/2013A&A...551A..63B} {551, A63}

\bibitem[\protect\citeauthoryear{{Bourrier}, {Ehrenreich}  \& {Lecavelier des Etangs}}{{Bourrier} et~al.}{2015}]{Bourrier2015}
{Bourrier} V.,  {Ehrenreich} D.,   {Lecavelier des Etangs} A.,  2015, \mn@doi [A\&A] {10.1051/0004-6361/201526894}, 582, A65

\bibitem[\protect\citeauthoryear{{Bourrier}, {Lecavelier des Etangs}, {Ehrenreich}, {Tanaka}  \& {Vidotto}}{{Bourrier} et~al.}{2016}]{Bourrier2016}
{Bourrier} V.,  {Lecavelier des Etangs} A.,  {Ehrenreich} D.,  {Tanaka} Y.~A.,   {Vidotto} A.~A.,  2016, \mn@doi [\aap] {10.1051/0004-6361/201628362}, \href {https://ui.adsabs.harvard.edu/abs/2016A&A...591A.121B} {591, A121}

\bibitem[\protect\citeauthoryear{{Bourrier}, {Ehrenreich}, {King}, {Lecavelier des Etangs}, {Wheatley}, {Vidal-Madjar}, {Pepe}  \& {Udry}}{{Bourrier} et~al.}{2017}]{Bourrier2017}
{Bourrier} V.,  {Ehrenreich} D.,  {King} G.,  {Lecavelier des Etangs} A.,  {Wheatley} P.~J.,  {Vidal-Madjar} A.,  {Pepe} F.,   {Udry} S.,  2017, \mn@doi [\aap] {10.1051/0004-6361/201629253}, \href {https://ui.adsabs.harvard.edu/abs/2017A&A...597A..26B} {597, A26}

\bibitem[\protect\citeauthoryear{{Bourrier} et~al.,}{{Bourrier} et~al.}{2022}]{Bourrier2022}
{Bourrier} V.,  et~al., 2022, \mn@doi [\aap] {10.1051/0004-6361/202142559}, \href {https://ui.adsabs.harvard.edu/abs/2022A&A...663A.160B} {663, A160}

\bibitem[\protect\citeauthoryear{Byrne \& Hindmarsh}{Byrne \& Hindmarsh}{1975}]{Byrne1975}
Byrne G.~D.,  Hindmarsh A.~C.,  1975, \mn@doi [ACM Trans. Math. Softw.] {10.1145/355626.355636}, 1, 71

\bibitem[\protect\citeauthoryear{{Carolan}, {Vidotto}, {Villarreal D'Angelo}  \& {Hazra}}{{Carolan} et~al.}{2021a}]{Carolan2021a}
{Carolan} S.,  {Vidotto} A.~A.,  {Villarreal D'Angelo} C.,   {Hazra} G.,  2021a, \mn@doi [\mnras] {10.1093/mnras/staa3431}, \href {https://ui.adsabs.harvard.edu/abs/2021MNRAS.500.3382C} {500, 3382}

\bibitem[\protect\citeauthoryear{{Carolan}, {Vidotto}, {Hazra}, {Villarreal D'Angelo}  \& {Kubyshkina}}{{Carolan} et~al.}{2021b}]{Carolan2021b}
{Carolan} S.,  {Vidotto} A.~A.,  {Hazra} G.,  {Villarreal D'Angelo} C.,   {Kubyshkina} D.,  2021b, \mn@doi [\mnras] {10.1093/mnras/stab2947}, \href {https://ui.adsabs.harvard.edu/abs/2021MNRAS.508.6001C} {508, 6001}

\bibitem[\protect\citeauthoryear{Carroll-Nellenback, Frank, Liu, Quillen, Blackman  \& Dobbs-Dixon}{Carroll-Nellenback et~al.}{2016}]{Carroll-Nellenback2016}
Carroll-Nellenback J.,  Frank A.,  Liu B.,  Quillen A.~C.,  Blackman E.~G.,   Dobbs-Dixon I.,  2016, \mn@doi [Monthly Notices of the Royal Astronomical Society] {10.1093/mnras/stw3307}, 466, 2458

\bibitem[\protect\citeauthoryear{{Chen} \& {Rogers}}{{Chen} \& {Rogers}}{2016}]{Chen2016}
{Chen} H.,  {Rogers} L.~A.,  2016, \mn@doi [\apj] {10.3847/0004-637X/831/2/180}, \href {https://ui.adsabs.harvard.edu/abs/2016ApJ...831..180C} {831, 180}

\bibitem[\protect\citeauthoryear{Clarke \& Alexander}{Clarke \& Alexander}{2016}]{Clarke2016}
Clarke C.~J.,  Alexander R.~D.,  2016, \mn@doi [Monthly Notices of the Royal Astronomical Society] {10.1093/mnras/stw1178}, 460, 3044

\bibitem[\protect\citeauthoryear{{Cranmer}}{{Cranmer}}{2004}]{Cranmer2004}
{Cranmer} S.~R.,  2004, \mn@doi [American Journal of Physics] {10.1119/1.1775242}, \href {https://ui.adsabs.harvard.edu/abs/2004AmJPh..72.1397C} {72, 1397}

\bibitem[\protect\citeauthoryear{{Debrecht}, {Carroll-Nellenback}, {Frank}, {Blackman}, {Fossati}, {McCann}  \& {Murray-Clay}}{{Debrecht} et~al.}{2020}]{Debrecht2020}
{Debrecht} A.,  {Carroll-Nellenback} J.,  {Frank} A.,  {Blackman} E.~G.,  {Fossati} L.,  {McCann} J.,   {Murray-Clay} R.,  2020, \mn@doi [\mnras] {10.1093/mnras/staa351}, \href {https://ui.adsabs.harvard.edu/abs/2020MNRAS.493.1292D} {493, 1292}

\bibitem[\protect\citeauthoryear{{Debrecht}, {Carroll-Nellenback}, {Frank}, {Blackman}, {Fossati}, {Murray-Clay}  \& {McCann}}{{Debrecht} et~al.}{2022}]{Debrecht2022}
{Debrecht} A.,  {Carroll-Nellenback} J.,  {Frank} A.,  {Blackman} E.~G.,  {Fossati} L.,  {Murray-Clay} R.,   {McCann} J.,  2022, \mn@doi [\mnras] {10.1093/mnras/stac112}, \href {https://ui.adsabs.harvard.edu/abs/2022MNRAS.tmp..128D} {}

\bibitem[\protect\citeauthoryear{{Dos Santos} et~al.,}{{Dos Santos} et~al.}{2022}]{DosSantos2022}
{Dos Santos} L.~A.,  et~al., 2022, \mn@doi [\aap] {10.1051/0004-6361/202142038}, \href {https://ui.adsabs.harvard.edu/abs/2022A&A...659A..62D} {659, A62}

\bibitem[\protect\citeauthoryear{{Ehrenreich} et~al.,}{{Ehrenreich} et~al.}{2015}]{Eherenreich2015}
{Ehrenreich} D.,  et~al., 2015, \mn@doi [\nat] {10.1038/nature14501}, \href {https://ui.adsabs.harvard.edu/abs/2015Natur.522..459E} {522, 459}

\bibitem[\protect\citeauthoryear{{Foreman-Mackey}, {Hogg}, {Lang}  \& {Goodman}}{{Foreman-Mackey} et~al.}{2013}]{ForemanMackey2013}
{Foreman-Mackey} D.,  {Hogg} D.~W.,  {Lang} D.,   {Goodman} J.,  2013, \mn@doi [\pasp] {10.1086/670067}, \href {https://ui.adsabs.harvard.edu/abs/2013PASP..125..306F} {125, 306}

\bibitem[\protect\citeauthoryear{{Fukue} \& {Okada}}{{Fukue} \& {Okada}}{1990}]{Fukue1990}
{Fukue} J.,  {Okada} R.,  1990, \pasj, \href {https://ui.adsabs.harvard.edu/abs/1990PASJ...42..249F} {42, 249}

\bibitem[\protect\citeauthoryear{Fulton et~al.,}{Fulton et~al.}{2017}]{Fulton2017}
Fulton B.~J.,  et~al., 2017, \mn@doi [The Astronomical Journal] {10.3847/1538-3881/aa80eb}, 154, 109

\bibitem[\protect\citeauthoryear{{Garc{\'\i}a Mu{\~n}oz}, {Youngblood}, {Fossati}, {Gandolfi}, {Cabrera}  \& {Rauer}}{{Garc{\'\i}a Mu{\~n}oz} et~al.}{2020}]{GarciaMunoz2020}
{Garc{\'\i}a Mu{\~n}oz} A.,  {Youngblood} A.,  {Fossati} L.,  {Gandolfi} D.,  {Cabrera} J.,   {Rauer} H.,  2020, \mn@doi [\apjl] {10.3847/2041-8213/ab61ff}, \href {https://ui.adsabs.harvard.edu/abs/2020ApJ...888L..21G} {888, L21}

\bibitem[\protect\citeauthoryear{{Gelman}, {Carlin}, {Stern}, {Dunson}, {Vehtari}  \& {Rubin}}{{Gelman} et~al.}{2014}]{Gelman2014}
{Gelman} A.,  {Carlin} J.~B.,  {Stern} H.~S.,  {Dunson} D.~B.,  {Vehtari} A.,   {Rubin} D.~B.,  2014, {Bayesian Data Analysis}

\bibitem[\protect\citeauthoryear{{Goodman} \& {Weare}}{{Goodman} \& {Weare}}{2010}]{Goodman2010}
{Goodman} J.,  {Weare} J.,  2010, \mn@doi [Communications in Applied Mathematics and Computational Science] {10.2140/camcos.2010.5.65}, \href {https://ui.adsabs.harvard.edu/abs/2010CAMCS...5...65G} {5, 65}

\bibitem[\protect\citeauthoryear{{Gordino} et~al.,}{{Gordino} et~al.}{2022}]{Gordino2022}
{Gordino} M.,  et~al., 2022, \mn@doi [\aap] {10.1051/0004-6361/202141960}, \href {https://ui.adsabs.harvard.edu/abs/2022A&A...657A..86G} {657, A86}

\bibitem[\protect\citeauthoryear{{Hazra}, {Vidotto}, {Carolan}, {Villarreal D'Angelo}  \& {Manchester}}{{Hazra} et~al.}{2022}]{Hazra2022}
{Hazra} G.,  {Vidotto} A.~A.,  {Carolan} S.,  {Villarreal D'Angelo} C.,   {Manchester} W.,  2022, \mn@doi [\mnras] {10.1093/mnras/stab3271}, \href {https://ui.adsabs.harvard.edu/abs/2022MNRAS.509.5858H} {509, 5858}

\bibitem[\protect\citeauthoryear{Holmstr{\"o}m, Ekenb{\"a}ck, Selsis, Penz, Lammer  \& Wurz}{Holmstr{\"o}m et~al.}{2008}]{Holmstrom2008}
Holmstr{\"o}m M.,  Ekenb{\"a}ck A.,  Selsis F.,  Penz T.,  Lammer H.,   Wurz P.,  2008, \mn@doi [Nature] {10.1038/nature06600}, 451, 970

\bibitem[\protect\citeauthoryear{{Izidoro}, {Schlichting}, {Isella}, {Dasgupta}, {Zimmermann}  \& {Bitsch}}{{Izidoro} et~al.}{2022}]{Izidoro2022}
{Izidoro} A.,  {Schlichting} H.~E.,  {Isella} A.,  {Dasgupta} R.,  {Zimmermann} C.,   {Bitsch} B.,  2022, \mn@doi [\apjl] {10.3847/2041-8213/ac990d}, \href {https://ui.adsabs.harvard.edu/abs/2022ApJ...939L..19I} {939, L19}

\bibitem[\protect\citeauthoryear{{Khodachenko}, {Shaikhislamov}, {Lammer}, {Berezutsky}, {Miroshnichenko}, {Rumenskikh}, {Kislyakova}  \& {Dwivedi}}{{Khodachenko} et~al.}{2019}]{Khodachenko2019}
{Khodachenko} M.~L.,  {Shaikhislamov} I.~F.,  {Lammer} H.,  {Berezutsky} A.~G.,  {Miroshnichenko} I.~B.,  {Rumenskikh} M.~S.,  {Kislyakova} K.~G.,   {Dwivedi} N.~K.,  2019, \mn@doi [\apj] {10.3847/1538-4357/ab46a4}, \href {https://ui.adsabs.harvard.edu/abs/2019ApJ...885...67K} {885, 67}

\bibitem[\protect\citeauthoryear{{Kirk}, {Alam}, {L{\'o}pez-Morales}  \& {Zeng}}{{Kirk} et~al.}{2020}]{Kirk2020}
{Kirk} J.,  {Alam} M.~K.,  {L{\'o}pez-Morales} M.,   {Zeng} L.,  2020, \mn@doi [\aj] {10.3847/1538-3881/ab6e66}, \href {https://ui.adsabs.harvard.edu/abs/2020AJ....159..115K} {159, 115}

\bibitem[\protect\citeauthoryear{{Kislyakova}, {Holmstr{\"o}m}, {Lammer}, {Odert}  \& {Khodachenko}}{{Kislyakova} et~al.}{2014}]{Kislyakova2014}
{Kislyakova} K.~G.,  {Holmstr{\"o}m} M.,  {Lammer} H.,  {Odert} P.,   {Khodachenko} M.~L.,  2014, \mn@doi [Science] {10.1126/science.1257829}, \href {https://ui.adsabs.harvard.edu/abs/2014Sci...346..981K} {346, 981}

\bibitem[\protect\citeauthoryear{Kramida, {Yu.~Ralchenko}, Reader  \& {and NIST ASD Team}}{Kramida et~al.}{2022}]{NIST_ASD}
Kramida A.,  {Yu.~Ralchenko} Reader J.,   {and NIST ASD Team} 2022, {NIST Atomic Spectra Database (ver. 5.10), [Online]. Available: {\tt{https://physics.nist.gov/asd}} [2017, April 9]. National Institute of Standards and Technology, Gaithersburg, MD.}

\bibitem[\protect\citeauthoryear{{Kulow}, {France}, {Linsky}  \& {Loyd}}{{Kulow} et~al.}{2014}]{Kulow2014}
{Kulow} J.~R.,  {France} K.,  {Linsky} J.,   {Loyd} R.~O.~P.,  2014, \mn@doi [\apj] {10.1088/0004-637X/786/2/132}, \href {https://ui.adsabs.harvard.edu/abs/2014ApJ...786..132K} {786, 132}

\bibitem[\protect\citeauthoryear{{Landsman} \& {Simon}}{{Landsman} \& {Simon}}{1993}]{Landsman1993}
{Landsman} W.,  {Simon} T.,  1993, \mn@doi [\apj] {10.1086/172589}, \href {https://ui.adsabs.harvard.edu/abs/1993ApJ...408..305L} {408, 305}

\bibitem[\protect\citeauthoryear{{Lavie} et~al.,}{{Lavie} et~al.}{2017}]{Lavie2017}
{Lavie} B.,  et~al., 2017, \mn@doi [\aap] {10.1051/0004-6361/201731340}, \href {https://ui.adsabs.harvard.edu/abs/2017A&A...605L...7L} {605, L7}

\bibitem[\protect\citeauthoryear{{Lecavelier Des Etangs} et~al.,}{{Lecavelier Des Etangs} et~al.}{2010}]{LecavelierDesEtangs2010}
{Lecavelier Des Etangs} A.,  et~al., 2010, \mn@doi [\aap] {10.1051/0004-6361/200913347}, \href {https://ui.adsabs.harvard.edu/abs/2010A&A...514A..72L} {514, A72}

\bibitem[\protect\citeauthoryear{{Lee} \& {Connors}}{{Lee} \& {Connors}}{2021}]{Lee2021}
{Lee} E.~J.,  {Connors} N.~J.,  2021, \mn@doi [\apj] {10.3847/1538-4357/abd6c7}, \href {https://ui.adsabs.harvard.edu/abs/2021ApJ...908...32L} {908, 32}

\bibitem[\protect\citeauthoryear{{Lee}, {Karalis}  \& {Thorngren}}{{Lee} et~al.}{2022}]{Lee2022}
{Lee} E.~J.,  {Karalis} A.,   {Thorngren} D.~P.,  2022, arXiv e-prints, \href {https://ui.adsabs.harvard.edu/abs/2022arXiv220109898L} {p. arXiv:2201.09898}

\bibitem[\protect\citeauthoryear{{Lopez} \& {Fortney}}{{Lopez} \& {Fortney}}{2013}]{Lopez2013}
{Lopez} E.~D.,  {Fortney} J.~J.,  2013, \mn@doi [\apj] {10.1088/0004-637X/776/1/2}, \href {https://ui.adsabs.harvard.edu/abs/2013ApJ...776....2L} {776, 2}

\bibitem[\protect\citeauthoryear{{Lundkvist} et~al.,}{{Lundkvist} et~al.}{2016}]{Lundkvist2016}
{Lundkvist} M.~S.,  et~al., 2016, \mn@doi [Nature Communications] {10.1038/ncomms11201}, \href {https://ui.adsabs.harvard.edu/abs/2016NatCo...711201L} {7, 11201}

\bibitem[\protect\citeauthoryear{{Luque} \& {Pall{\'e}}}{{Luque} \& {Pall{\'e}}}{2022}]{Luque2022}
{Luque} R.,  {Pall{\'e}} E.,  2022, \mn@doi [Science] {10.1126/science.abl7164}, \href {https://ui.adsabs.harvard.edu/abs/2022Sci...377.1211L} {377, 1211}

\bibitem[\protect\citeauthoryear{{MacLeod} \& {Oklop{\v{c}}i{\'c}}}{{MacLeod} \& {Oklop{\v{c}}i{\'c}}}{2022}]{Macleod2022}
{MacLeod} M.,  {Oklop{\v{c}}i{\'c}} A.,  2022, \mn@doi [\apj] {10.3847/1538-4357/ac46ce}, \href {https://ui.adsabs.harvard.edu/abs/2022ApJ...926..226M} {926, 226}

\bibitem[\protect\citeauthoryear{{Matsakos}, {Uribe}  \& {K{\"o}nigl}}{{Matsakos} et~al.}{2015}]{Matsakos2015}
{Matsakos} T.,  {Uribe} A.,   {K{\"o}nigl} A.,  2015, \mn@doi [\aap] {10.1051/0004-6361/201425593}, \href {https://ui.adsabs.harvard.edu/abs/2015A&A...578A...6M} {578, A6}

\bibitem[\protect\citeauthoryear{{McCann}, {Murray-Clay}, {Kratter}  \& {Krumholz}}{{McCann} et~al.}{2019}]{McCann2019}
{McCann} J.,  {Murray-Clay} R.~A.,  {Kratter} K.,   {Krumholz} M.~R.,  2019, \mn@doi [\apj] {10.3847/1538-4357/ab05b8}, \href {https://ui.adsabs.harvard.edu/abs/2019ApJ...873...89M} {873, 89}

\bibitem[\protect\citeauthoryear{{Murray-Clay}, {Chiang}  \& {Murray}}{{Murray-Clay} et~al.}{2009}]{Murray-Clay2009}
{Murray-Clay} R.~A.,  {Chiang} E.~I.,   {Murray} N.,  2009, \mn@doi [\apj] {10.1088/0004-637X/693/1/23}, \href {https://ui.adsabs.harvard.edu/abs/2009ApJ...693...23M} {693, 23}

\bibitem[\protect\citeauthoryear{{Oklop{\v{c}}i{\'c}}}{{Oklop{\v{c}}i{\'c}}}{2019}]{Oklopvcic2019}
{Oklop{\v{c}}i{\'c}} A.,  2019, \mn@doi [\apj] {10.3847/1538-4357/ab2f7f}, \href {https://ui.adsabs.harvard.edu/abs/2019ApJ...881..133O} {881, 133}

\bibitem[\protect\citeauthoryear{{Owen} \& {Alvarez}}{{Owen} \& {Alvarez}}{2016}]{Owen2016}
{Owen} J.~E.,  {Alvarez} M.~A.,  2016, \mn@doi [\apj] {10.3847/0004-637X/816/1/34}, \href {https://ui.adsabs.harvard.edu/abs/2016ApJ...816...34O} {816, 34}

\bibitem[\protect\citeauthoryear{{Owen} \& {Jackson}}{{Owen} \& {Jackson}}{2012}]{Owen2012}
{Owen} J.~E.,  {Jackson} A.~P.,  2012, \mn@doi [\mnras] {10.1111/j.1365-2966.2012.21481.x}, \href {https://ui.adsabs.harvard.edu/abs/2012MNRAS.425.2931O} {425, 2931}

\bibitem[\protect\citeauthoryear{{Owen} \& {Schlichting}}{{Owen} \& {Schlichting}}{2024}]{Owen2024}
{Owen} J.~E.,  {Schlichting} H.~E.,  2024, \mn@doi [\mnras] {10.1093/mnras/stad3972}, \href {https://ui.adsabs.harvard.edu/abs/2024MNRAS.528.1615O} {528, 1615}

\bibitem[\protect\citeauthoryear{{Owen} \& {Wu}}{{Owen} \& {Wu}}{2013}]{Owen2013}
{Owen} J.~E.,  {Wu} Y.,  2013, \mn@doi [\apj] {10.1088/0004-637X/775/2/105}, \href {https://ui.adsabs.harvard.edu/abs/2013ApJ...775..105O} {775, 105}

\bibitem[\protect\citeauthoryear{{Owen} et~al.,}{{Owen} et~al.}{2023}]{Owen2023}
{Owen} J.~E.,  et~al., 2023, \mn@doi [\mnras] {10.1093/mnras/stac3414}, \href {https://ui.adsabs.harvard.edu/abs/2023MNRAS.518.4357O} {518, 4357}

\bibitem[\protect\citeauthoryear{{Parker}}{{Parker}}{1958}]{Parker1958}
{Parker} E.~N.,  1958, \mn@doi [\apj] {10.1086/146579}, \href {https://ui.adsabs.harvard.edu/abs/1958ApJ...128..664P} {128, 664}

\bibitem[\protect\citeauthoryear{Petzold}{Petzold}{1983}]{Petzold1983}
Petzold L.,  1983, \mn@doi [SIAM Journal on Scientific and Statistical Computing] {10.1137/0904010}, 4, 136

\bibitem[\protect\citeauthoryear{{Pinto} \& {Galli}}{{Pinto} \& {Galli}}{2008}]{Pinto2008}
{Pinto} C.,  {Galli} D.,  2008, \mn@doi [\aap] {10.1051/0004-6361:20078819}, \href {https://ui.adsabs.harvard.edu/abs/2008A&A...484...17P} {484, 17}

\bibitem[\protect\citeauthoryear{{Raga}, {Cabrit}  \& {Canto}}{{Raga} et~al.}{1995}]{Raga1995}
{Raga} A.~C.,  {Cabrit} S.,   {Canto} J.,  1995, \mn@doi [\mnras] {10.1093/mnras/273.2.422}, \href {https://ui.adsabs.harvard.edu/abs/1995MNRAS.273..422R} {273, 422}

\bibitem[\protect\citeauthoryear{Rockcliffe et~al.,}{Rockcliffe et~al.}{2021}]{Rockcliffe_2021}
Rockcliffe K.~E.,  et~al., 2021, \mn@doi [The Astronomical Journal] {10.3847/1538-3881/ac126f}, 162, 116

\bibitem[\protect\citeauthoryear{{Rockcliffe}, {Newton}, {Youngblood}, {Duvvuri}, {Plavchan}, {Gao}, {Mann}  \& {Lowrance}}{{Rockcliffe} et~al.}{2023}]{Rockcliffe2023}
{Rockcliffe} K.~E.,  {Newton} E.~R.,  {Youngblood} A.,  {Duvvuri} G.~M.,  {Plavchan} P.,  {Gao} P.,  {Mann} A.~W.,   {Lowrance} P.~J.,  2023, \mn@doi [\aj] {10.3847/1538-3881/ace536}, \href {https://ui.adsabs.harvard.edu/abs/2023AJ....166...77R} {166, 77}

\bibitem[\protect\citeauthoryear{{Schreyer}, {Owen}, {Spake}, {Bahroloom}  \& {Di Giampasquale}}{{Schreyer} et~al.}{2024}]{Schreyer2024}
{Schreyer} E.,  {Owen} J.~E.,  {Spake} J.~J.,  {Bahroloom} Z.,   {Di Giampasquale} S.,  2024, \mn@doi [\mnras] {10.1093/mnras/stad3528}, \href {https://ui.adsabs.harvard.edu/abs/2024MNRAS.527.5117S} {527, 5117}

\bibitem[\protect\citeauthoryear{{Schulik} \& {Booth}}{{Schulik} \& {Booth}}{2023}]{Schulik2023}
{Schulik} M.,  {Booth} R.~A.,  2023, \mn@doi [\mnras] {10.1093/mnras/stad1251}, \href {https://ui.adsabs.harvard.edu/abs/2023MNRAS.523..286S} {523, 286}

\bibitem[\protect\citeauthoryear{{Schulreich} \& {Breitschwerdt}}{{Schulreich} \& {Breitschwerdt}}{2011}]{Schulreich2011}
{Schulreich} M.~M.,  {Breitschwerdt} D.,  2011, \mn@doi [\aap] {10.1051/0004-6361/200912436}, \href {https://ui.adsabs.harvard.edu/abs/2011A&A...531A..13S} {531, A13}

\bibitem[\protect\citeauthoryear{{Schultz}, {Krstic}, {Lee}  \& {Raymond}}{{Schultz} et~al.}{2008}]{Schultz2008}
{Schultz} D.~R.,  {Krstic} P.~S.,  {Lee} T.~G.,   {Raymond} J.~C.,  2008, \mn@doi [\apj] {10.1086/533579}, \href {https://ui.adsabs.harvard.edu/abs/2008ApJ...678..950S} {678, 950}

\bibitem[\protect\citeauthoryear{{Schunk} \& {Nagy}}{{Schunk} \& {Nagy}}{1980}]{Schunk1980}
{Schunk} R.~W.,  {Nagy} A.~F.,  1980, \mn@doi [Reviews of Geophysics and Space Physics] {10.1029/RG018i004p00813}, \href {https://ui.adsabs.harvard.edu/abs/1980RvGSP..18..813S} {18, 813}

\bibitem[\protect\citeauthoryear{Shaikhislamov, Khodachenko, Lammer, Berezutsky, Miroshnichenko  \& Rumenskikh}{Shaikhislamov et~al.}{2020}]{Shaikhislamov2020}
Shaikhislamov I.~F.,  Khodachenko M.~L.,  Lammer H.,  Berezutsky A.~G.,  Miroshnichenko I.~B.,   Rumenskikh M.~S.,  2020, \mn@doi [Monthly Notices of the Royal Astronomical Society] {10.1093/mnras/staa2367}, 500, 1404

\bibitem[\protect\citeauthoryear{{Sing} et~al.,}{{Sing} et~al.}{2019}]{Sing2019}
{Sing} D.~K.,  et~al., 2019, \mn@doi [\aj] {10.3847/1538-3881/ab2986}, \href {https://ui.adsabs.harvard.edu/abs/2019AJ....158...91S} {158, 91}

\bibitem[\protect\citeauthoryear{{Spake} et~al.,}{{Spake} et~al.}{2018}]{Spake2018}
{Spake} J.~J.,  et~al., 2018, \mn@doi [\nat] {10.1038/s41586-018-0067-5}, \href {https://ui.adsabs.harvard.edu/abs/2018Natur.557...68S} {557, 68}

\bibitem[\protect\citeauthoryear{{Spitzer}}{{Spitzer}}{1978}]{Spitzer1978}
{Spitzer} L.,  1978, {Physical processes in the interstellar medium}, \mn@doi{10.1002/9783527617722.
}

\bibitem[\protect\citeauthoryear{Szab{\'{o}} \& Kiss}{Szab{\'{o}} \& Kiss}{2011}]{Szabo2011}
Szab{\'{o}} G.~M.,  Kiss L.~L.,  2011, \mn@doi [The Astrophysical Journal] {10.1088/2041-8205/727/2/l44}, 727, L44

\bibitem[\protect\citeauthoryear{{Tanaka}, {Suzuki}  \& {Inutsuka}}{{Tanaka} et~al.}{2014}]{Tanaka2014}
{Tanaka} Y.~A.,  {Suzuki} T.~K.,   {Inutsuka} S.-i.,  2014, \mn@doi [\apj] {10.1088/0004-637X/792/1/18}, \href {https://ui.adsabs.harvard.edu/abs/2014ApJ...792...18T} {792, 18}

\bibitem[\protect\citeauthoryear{{Trammell}, {Arras}  \& {Li}}{{Trammell} et~al.}{2011}]{Trammell2011}
{Trammell} G.~B.,  {Arras} P.,   {Li} Z.-Y.,  2011, \mn@doi [\apj] {10.1088/0004-637X/728/2/152}, \href {https://ui.adsabs.harvard.edu/abs/2011ApJ...728..152T} {728, 152}

\bibitem[\protect\citeauthoryear{Tremblin \& Chiang}{Tremblin \& Chiang}{2012}]{Tremblin2012}
Tremblin P.,  Chiang E.,  2012, \mn@doi [Monthly Notices of the Royal Astronomical Society] {10.1093/mnras/sts212}, 428, 2565

\bibitem[\protect\citeauthoryear{{Valencia}, {Ikoma}, {Guillot}  \& {Nettelmann}}{{Valencia} et~al.}{2010}]{Valencia2010}
{Valencia} D.,  {Ikoma} M.,  {Guillot} T.,   {Nettelmann} N.,  2010, \mn@doi [\aap] {10.1051/0004-6361/200912839}, \href {https://ui.adsabs.harvard.edu/abs/2010A&A...516A..20V} {516, A20}

\bibitem[\protect\citeauthoryear{Van~Eylen, Agentoft, Lundkvist, Kjeldsen, Owen, Fulton, Petigura  \& Snellen}{Van~Eylen et~al.}{2018}]{VanEylen2018}
Van~Eylen V.,  Agentoft C.,  Lundkvist M.~S.,  Kjeldsen H.,  Owen J.~E.,  Fulton B.~J.,  Petigura E.,   Snellen I.,  2018, \mn@doi [Monthly Notices of the Royal Astronomical Society] {10.1093/mnras/sty1783}, 479, 4786

\bibitem[\protect\citeauthoryear{{Venturini}, {Guilera}, {Haldemann}, {Ronco}  \& {Mordasini}}{{Venturini} et~al.}{2020}]{Venturini2020}
{Venturini} J.,  {Guilera} O.~M.,  {Haldemann} J.,  {Ronco} M.~P.,   {Mordasini} C.,  2020, \mn@doi [\aap] {10.1051/0004-6361/202039141}, \href {https://ui.adsabs.harvard.edu/abs/2020A&A...643L...1V} {643, L1}

\bibitem[\protect\citeauthoryear{{Vidal-Madjar}, {Lecavelier des Etangs}, {D{\'e}sert}, {Ballester}, {Ferlet}, {H{\'e}brard}  \& {Mayor}}{{Vidal-Madjar} et~al.}{2003}]{Vidal-Madjar2003}
{Vidal-Madjar} A.,  {Lecavelier des Etangs} A.,  {D{\'e}sert} J.~M.,  {Ballester} G.~E.,  {Ferlet} R.,  {H{\'e}brard} G.,   {Mayor} M.,  2003, \mn@doi [\nat] {10.1038/nature01448}, \href {https://ui.adsabs.harvard.edu/abs/2003Natur.422..143V} {422, 143}

\bibitem[\protect\citeauthoryear{{Vidal-Madjar} et~al.,}{{Vidal-Madjar} et~al.}{2004}]{Vidal-Madjar2004}
{Vidal-Madjar} A.,  et~al., 2004, \mn@doi [\apjl] {10.1086/383347}, \href {https://ui.adsabs.harvard.edu/abs/2004ApJ...604L..69V} {604, L69}

\bibitem[\protect\citeauthoryear{{Villarreal D'Angelo}, {Vidotto}, {Esquivel}, {Hazra}  \& {Youngblood}}{{Villarreal D'Angelo} et~al.}{2021}]{Villarreal2021}
{Villarreal D'Angelo} C.,  {Vidotto} A.~A.,  {Esquivel} A.,  {Hazra} G.,   {Youngblood} A.,  2021, \mn@doi [\mnras] {10.1093/mnras/staa3867}, \href {https://ui.adsabs.harvard.edu/abs/2021MNRAS.501.4383V} {501, 4383}

\bibitem[\protect\citeauthoryear{{Virtanen} et~al.,}{{Virtanen} et~al.}{2020}]{scipy}
{Virtanen} P.,  et~al., 2020, \mn@doi [Nature Methods] {10.1038/s41592-019-0686-2}, \href {https://ui.adsabs.harvard.edu/abs/2020NatMe..17..261V} {17, 261}

\bibitem[\protect\citeauthoryear{Waugh}{Waugh}{1961}]{Waugh1961}
Waugh 1961, Technical report, Evaluation Of Integral Of Elliptic Gaussian Distribution Over A Centred Ellipse, \url {https://apps.dtic.mil/sti/pdfs/AD0271430.pdf}.
U.S Navy, \url {https://apps.dtic.mil/sti/pdfs/AD0271430.pdf}

\bibitem[\protect\citeauthoryear{{Wilson} et~al.,}{{Wilson} et~al.}{2022}]{wilson2022}
{Wilson} D.~J.,  et~al., 2022, \mn@doi [\apj] {10.3847/1538-4357/ac87a8}, \href {https://ui.adsabs.harvard.edu/abs/2022ApJ...936..189W} {936, 189}

\bibitem[\protect\citeauthoryear{{Zeng} et~al.,}{{Zeng} et~al.}{2019}]{Zeng2019}
{Zeng} L.,  et~al., 2019, \mn@doi [Proceedings of the National Academy of Science] {10.1073/pnas.1812905116}, \href {https://ui.adsabs.harvard.edu/abs/2019PNAS..116.9723Z} {116, 9723}

\bibitem[\protect\citeauthoryear{{Zhang} et~al.,}{{Zhang} et~al.}{2022}]{Zhang2022}
{Zhang} M.,  et~al., 2022, \mn@doi [\aj] {10.3847/1538-3881/ac3f3b}, \href {https://ui.adsabs.harvard.edu/abs/2022AJ....163...68Z} {163, 68}

\bibitem[\protect\citeauthoryear{{Zhang}, {Knutson}, {Dai}, {Wang}, {Ricker}, {Schwarz}, {Mann}  \& {Collins}}{{Zhang} et~al.}{2023}]{ZhangM2023}
{Zhang} M.,  {Knutson} H.~A.,  {Dai} F.,  {Wang} L.,  {Ricker} G.~R.,  {Schwarz} R.~P.,  {Mann} C.,   {Collins} K.,  2023, \mn@doi [\aj] {10.3847/1538-3881/aca75b}, \href {https://ui.adsabs.harvard.edu/abs/2023AJ....165...62Z} {165, 62}

\bibitem[\protect\citeauthoryear{{dos Santos} et~al.,}{{dos Santos} et~al.}{2019}]{dosSantos2019}
{dos Santos} L.~A.,  et~al., 2019, \mn@doi [\aap] {10.1051/0004-6361/201935663}, \href {https://ui.adsabs.harvard.edu/abs/2019A&A...629A..47D} {629, A47}

\bibitem[\protect\citeauthoryear{{dos Santos} et~al.,}{{dos Santos} et~al.}{2020}]{dosSantos2020}
{dos Santos} L.~A.,  et~al., 2020, \mn@doi [\aap] {10.1051/0004-6361/201937327}, \href {https://ui.adsabs.harvard.edu/abs/2020A&A...634L...4D} {634, L4}

\makeatother
\end{thebibliography}




\appendix

\section{Mapping the optically thin ionization rate to an EUV luminosity}
\label{appendix2}

In order to map the optically thin ionization rate at the planet to an integrated EUV luminosity of the star, one needs a stellar spectrum. As the interstellar medium extinguishes much EUV radiation from stars, both the integrated EUV luminosity of planet-hosting stars and the shape of the spectrum have significant uncertainties. This means that the range of optically thin photoionization rates accounts for variations in both these factors.    

Assuming we have a "perfect" stellar spectrum, $f_\nu$, that has been normalised to unity in the EUV range, we can map the integrated EUV luminosity to the optically thin ionization rate as follows: 

\begin{align}
    \Gamma = L_{\text{EUV}}\int_{13.6~{\rm eV}}^{100~{\rm eV}}\frac{f_{\nu}}{4\pi r^2h\nu}\sigma_{\nu}d\nu
\end{align}
where $L_{\text{EUV}}$ is the luminosity integrated over the EUV range. As $\sigma_\nu$ peaks at $13.6$ eV and decreases strongly as a function as a function of frequency, $\propto \left(\frac{\bar{\nu}}{\nu}\right)^3$ \citep[e.g.,][]{Spitzer1978}, harder spectra lead to a smaller photoionization rate for the same integrated EUV luminosity.  

\section{Posterior Plots}
\label{appendix1}

\begin{figure*}
\centering
\includegraphics[width=\textwidth]{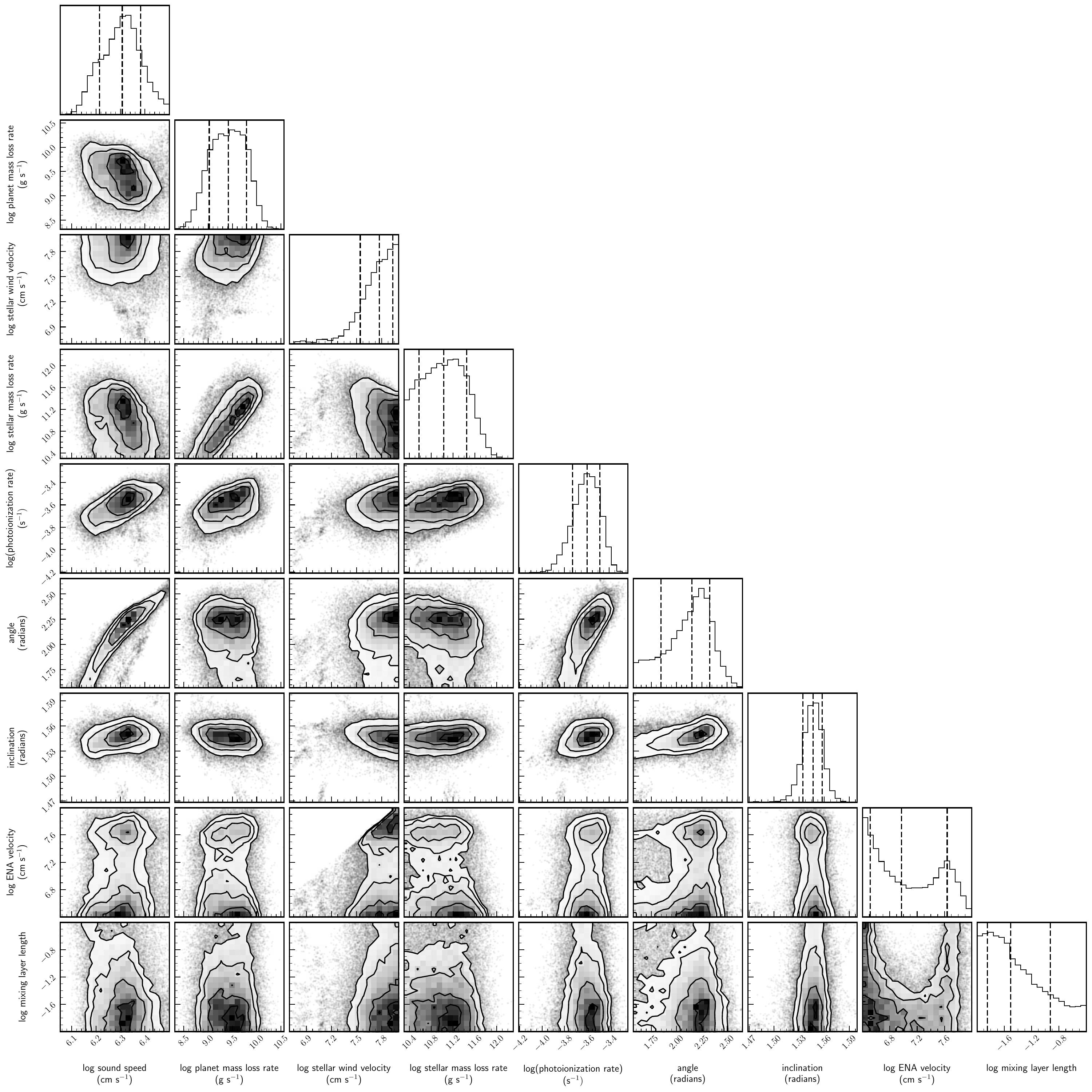}
\caption{The full posterior plot for our model shown in Section~\ref{sec:results}.}
\label{fig:full_posterior4}
\end{figure*}

We provide the full posterior plots for the model outlined in Section~\ref{sec:results} in Figure~\ref{fig:full_posterior4}.

\bsp	
\label{lastpage}
\end{document}